\documentstyle[preprint,aps,epsf]{revtex}
\begin{document}

\draft

\title{Non-Fermi liquid theory of a compactified Anderson single-impurity model}

\author{Guang-Ming Zhang and A. C. Hewson}

\address{Department of Mathematics, Imperial College, London SW7 2BZ, England.}

\maketitle

\begin{abstract}
{ We consider a version of the symmetric Anderson
impurity model (compactified) which has a non-Fermi liquid weak
 coupling regime. We find that in the Majorana fermion representation  
the perturbation theory can be 
conveniently developed in terms of Pfaffian determinants and we use this
 formalism to calculate  the impurity free energy, self energies,
 and vertex functions. We also derive expressions for the 
impurity and the local conduction electron charge and spin dynamical 
susceptibilities in terms of the impurity self-energies and 
vertex functions. In the second-order perturbation theory, a linear temperature 
dependence of the electrical resistivity is obtained, and the leading 
corrections  to 
the impurity specific heat are found to  behave as $T\ln T$.  The impurity 
static susceptibilities have terms in $\ln T$ to zero, first, and second order,
and corrections of $\ln^2 T$ to second order as well. The conduction 
electron static susceptibilities, and the singlet superconducting paired static
susceptibility at the impurity site, are found to have  second-order 
corrections $\ln T$, which we interpret as an indication that  a singlet 
conduction electron pairing resonance forms at the Fermi level (the chemical 
potential). When the perturbation theory is extended to third 
order logarithmic divergences are found in the only vertex function 
$\Gamma_{0,1,2,3}(0,0,0,0)$, which is non-vanishing in the zero frequency 
limit.  We use  the multiplicative renormalization-group 
(RG) method to sum  all the leading order logarithmic contributions. These 
give rise to a new weak-coupling low-temperature energy scale
 $T_c=\Delta{\rm exp}
\left [- \frac{1}{9} \left ( \frac{\pi\Delta}{U} \right )^{2} \right ]$,
which is the combination of the two independent coupling parameters. 
The RG scaling equation is also derived and shows that the 
dimensionless coupling constant $\bar{U}=\frac{U}{\pi\Delta}$ is increased  
as the high-energy scale $\Delta$ is reduced, so our perturbational 
results can be justified in the regime $T>T_c$. Below $T_c$ the 
perturbation theory breaks down.}
\end{abstract}

\pacs{75.15.Qm,71.45.-d,75.15.Nj}

\section{Introduction}

Strongly correlated electron systems, and especially the high ${\rm T_c}$ 
cuprate superconductors, display features which have been difficult to 
reconcile with conventional theories of Fermi liquids and  superconductivity.  
The various ideas that have been put forward to explain the behavior of these 
systems remain controversial. The unusual normal state of the high ${\rm T_c}$ 
materials  has been interpreted as some form of non-Fermi liquid (non-FL), 
but there have been no fully convincing microscopic derivations of such a 
state in two dimensional systems with short range interactions. Understanding 
the normal state is rather important as it may be
the key to understanding the nature and origin of  the superconductivity. 
Anomalous behavior that has been observed experimentally in certain heavy 
fermion U-based 
superconductors has also been interpreted as some form of non-FL behavior
 \cite{maple}. One of the most striking characteristics of the non-FL in both 
these types of systems is the 
linear temperature dependence of the electrical resistivity.
A phenomenological marginal-FL spectrum for the spin and charge
fluctuations was put forward by Varma {\it et al.} \cite{varma}
as a unified way of interpreting the diverse anomalies observed in 
the cuprate superconductors. The essential
point in this theory is that the frequency dependence of the polarization is 
singular, while the momentum dependence is taken to be smooth. This picture is 
very similar  to  the two dimensional Luttinger liquid hypothesis used by  
Anderson \cite{and}. There is no generally accepted microscopic  theory to 
explain the linear temperature dependence of the  electrical resistivity.

The lack of progress in developing a  microscopic theory is due to the 
difficulty in solving the  lattice models in strong coupling, and it
may be quite a time before suitable methods can be developed to overcome
these difficulties.
There are  impurity models, however, which display non-FL
behavior; these are more accessible, and may provide valuable insights.
For many of these models we have exact solutions \cite{hewson}, and in this 
context, the non-FL thermodynamic behavior of the 
two-channel Kondo model has been extensively investigated 
\cite{nb,adtw,al,ek,cit,cs}. However the exact 
solution for this model obtained from conformal field theory \cite{al} gives a 
resistivity that behaves as $\rho(T)=\rho(0)(1+a\sqrt{T})$ in the low 
temperature limit, so it is not clear that  the experimental observations can 
be interpreted using this theoretical model. The question then arises  as to 
whether the linear temperature dependence of the  resistivity, and  the 
marginal-FL behavior, can be found in any single-impurity model.

In this paper we examine a model, originally put forward by Coleman and 
Schofield \cite{cs}, which we will show displays just this type of 
behavior, at least in weak coupling perturbation theory.
We can obtain the model starting from  usual symmetric Anderson single-impurity 
model in the Majorana representation. The usual model has O(4) symmetry arising 
from a product of the SU(2) spin and the SU(2) particle-hole symmetry.
The modified model is obtained by reducing the  O(4) symmetry of the 
hybridization term to O(3) symmetry.  We then construct a systematic 
non-FL perturbation theory 
around the weak coupling limit. In contrast to the two-channel Kondo model, the 
linear temperature dependence of the electrical resistivity is obtained from
second order perturbation theory. The result would appear to be valid at 
temperatures above a characteristic
temperature $T_c$. At $T_c$,  the sum of the leading order 
singular perturbational contributions to the electrical resistivity diverge
indicating a breakdown of  the perturbation theory at this energy scale. 
A brief report of this work was given in our previous paper \cite{zhang}.  

The paper is arranged as follows. In section II, the modified Anderson 
single-impurity model is introduced and the general features of this model are 
discussed, in particular, there is an important relation between the conduction
electron and impurity Green functions. In section III, we fully investigate the 
unperturbed Hamiltonian, and show that it displays a non-FL behavior in the 
spin and charge 
density-density spectrum. In section IV, we present the a new 
perturbation formalism in which the impurity free energy, single-particle Green 
functions, and two-particle vertex functions can be expressed in terms of 
Pfaffian determinants, a specific determinant defined from the ordinary 
antisymmetric determinant. In section V, using this formalism, expressions are 
derived for both  impurity and local conduction electron
spin and charge dynamical susceptibilities, including the conduction electron 
singlet superconducting paired susceptibility, in terms of the 
impurity self energy and vertex functions. In section VI, the lower order 
results are given and discussed in detail. In section VII, multiplicative 
renormalization-group method is applied to this model to sum all the leading
logarithmic contributions and obtain the scaling equation. Finally, some 
discussions and conclusions are given in section VIII. 
  
\section{The model Hamiltonian}

The ordinary symmetric Anderson single-impurity model can be expressed in 
the form:
\begin{eqnarray}
&& H=it\sum_{n,\sigma}\left [ C^{\dag}_{\sigma}(n+1)C_{\sigma}(n)- H.c. \right ]
  +iV\sum_{\sigma}\left [C^{\dag}_{\sigma}(0)d_{\sigma}- H.c.\right ]
 \nonumber \\ && \hspace{2cm}
 +U \left(d^{\dag}_{\uparrow}d_{\uparrow}-\frac{1}{2}\right)
    \left(d^{\dag}_{\downarrow}d_{\downarrow}-\frac{1}{2}\right),
\end{eqnarray}
where the symmetric condition $\epsilon_d=-U/2$ has been used, and the chemical
potential is set to zero, the Fermi level. The Hamiltonian has $O(4)$ symmetry 
due to the $SU(2)$ symmetry from the spin rotational invariance and an 
additional $SU(2)$ from particle-hole symmetry, giving 
$O(4)\sim SU(2)\otimes SU(2)$.

It is important to observe that the $O(4)$ symmetry can be explicitly 
displayed when the fermions of each type of spin are expressed in terms of four
Majorana (real) fermions \cite{cs}:
$$ C_{\uparrow}(n)=\frac{1}{\sqrt{2}}\left [\Psi_1(n)-i\Psi_2(n) \right ],
  \hspace{1cm}
  C_{\downarrow}(n)=\frac{1}{\sqrt{2}}\left [-\Psi_3(n)-i\Psi_0(n)\right ] ; $$
$$ d_{\uparrow}=\frac{1}{\sqrt{2}}(d_1-id_2), \hspace{1cm}
   d_{\downarrow}=\frac{1}{\sqrt{2}}(-d_3-id_0), $$
these new operators satisfy 
\begin{eqnarray}
&& \Psi_{\alpha}(n)=(\Psi_{\alpha}(n))^{\dag}, 
   \hspace{1cm} d_{\alpha}=(d_{\alpha})^{\dag};
\nonumber \\ 
&& \{\Psi_{\alpha}(n), \Psi_{\beta}(n')\}
 =\delta_{\alpha,\beta}\delta_{n,n'},
  \hspace{1cm}  
 \{d_{\alpha}, d_{\beta}\} = \delta_{\alpha,\beta}.
\end{eqnarray}
Then the Hamiltonian becomes
\begin{equation}
H = it\sum_n\sum_{\alpha=0}^3\Psi_{\alpha}(n+1)\Psi_{\alpha}(n)
  +iV \sum_{\alpha=0}^3 \Psi_{\alpha}(0)d_{\alpha}+ Ud_1d_2d_3d_0.
\end{equation}
  
When the $O(4)$ symmetry breaks down to $O(3)$ 
symmetry in the hybridization term, the model Hamiltonian becomes
\begin{eqnarray}
&&  H=it\sum_n\sum_{\alpha=0}^3\Psi_{\alpha}(n+1)\Psi_{\alpha}(n)
 +iV_0\Psi_0(0)d_0 \nonumber \\ && \hspace{1cm}
  +iV\sum_{\alpha=1}^3\Psi_{\alpha}(0)d_{\alpha}+ Ud_1d_2d_3d_0,
\end{eqnarray}
where $V_0\neq V$. In the large $U$ limit ($U>0$), a Schrieffer-Wolff canonical
transformation can be applied to generate a s-d type of model, the so-called
`compactified two-channel Kondo model', 
\begin{equation}
H=it\sum_n\sum_{\alpha=0}^3\Psi_{\alpha}(n+1)\Psi_{\alpha}(n)
  +\left [J_1\vec{\sigma}(0)+J_2\vec{\tau}(0)\right] \cdot \vec{S}_d,
\end{equation}
with $J_1={2V(V+V_0)}/{U}$ and $J_2={2V(V-V_0)}/{U}$, where the local 
impurity spin couples to both the conduction electron spin $\vec{\sigma}(0)$ and
the conduction electron `isospin' density operator $\vec{\tau}(0)$ at
the impurity site. When $V_0=0$ the two exchange couplings are identical,
$i.e.$, $J_1=J_2$, and it had been conjectured that this form of the model has 
the same low-energy excitations as the two-channel Kondo model \cite{cit}.

To distinguish the Anderson model of Eq.(4) from other Anderson impurity models
we will refer to it as the `compactified' Anderson impurity model. Here we 
concentrate on the $V_0=0$ case: 
\begin{eqnarray}
 && H=H_0+H_I, \nonumber \\ 
 && H_0=it\sum_n\sum_{\alpha=0}^3\Psi_{\alpha}(n+1)\Psi_{\alpha}(n)
        +iV\sum_{\alpha=1}^3\Psi_{\alpha}(0)d_{\alpha},
  \nonumber \\ 
 && H_I = Ud_1d_2d_3d_0.
\end{eqnarray}
This will be used as our model Hamiltonian. The Hamiltonian for $V_0\ne0$
is considered in a separate paper \cite{hews}.

Fourier transforms for the conduction electron operators can be 
introduced as usual 
\begin{equation}
 \Psi_{\alpha}(n)=\frac{1}{\sqrt{N}}\sum_k\Psi_{\alpha}(k)e^{ikna}, 
  \hspace{1cm} \alpha=0,1,2,3,
\end{equation}
here $N$ is the total number of the sites and the lattice spacing is $a$. The
anticommutation relation for the conduction electrons is converted into
\begin{equation}
 \{ \Psi_{\alpha}(k),\Psi_{\beta}(-k')\} =\delta_{\alpha,\beta}\delta_{k,k'}.
\end{equation} 
Substituting these expressions in our model Hamiltonian, up to a constant we get
\begin{equation}
H=\sum_{k>0}\sum_{\alpha=0}^3\epsilon_k\Psi_{\alpha}(-k)\Psi_{\alpha}(k)
  +\frac{iV}{\sqrt{N}}\sum_k\sum_{\alpha=1}^3\Psi_{\alpha}(k)d_{\alpha}
  +Ud_1d_2d_3d_0,
\end{equation}
where $\epsilon_k=2t\sin(ka)$ is the dispersion relation of the conduction 
electrons. 

In the present model, the new O(3) symmetry in the  hybridization is the key 
feature. Since the scalar field ($\alpha=0$) decouples from the local impurity,
its single-particle Green functions, defined in terms of retarded 
double-time correlation function, is easily found to be a free propagator:
\begin{equation}
 G^0_{k,k'}(\omega_n) \equiv -\int^{\beta}_0 d\tau 
 \langle T_{\tau}\Psi_0(k,\tau)\Psi_0(-k',0)\rangle e^{i\omega_n\tau}
  = \frac{\delta_{k,k'}}{i\omega_n-\epsilon_k},
\end{equation}
where $\omega_n=(2n+1)\pi/\beta$, $\beta$ is the inverse of the temperature. 
Meanwhile, the vector field $\Psi_{\alpha}(k)$ ($\alpha=1,2,3$)  
hybridizes with the corresponding impurity vector field $d_{\alpha}$, and the 
scattering of the conduction electrons from the 
local impurity is given by the following relation:
\begin{equation}
 G^{\alpha}_{k,k'}(\omega_n)=\frac{\delta_{k,k'}}{i\omega_n-\epsilon_k}
  +\frac{V^2}{N}\frac{G_{\rm vec}(\omega_n)}
                     {(i\omega_n-\epsilon_k)(i\omega_n-\epsilon_{-k'})},
\end{equation}
where $G_{\rm vec}(\omega_n)$ is the Fourier transform of the impurity 
vector propagator defined by 
$-\langle T_{\tau}d_{\alpha}(\tau)d_{\alpha}(\tau')\rangle_H$. Then the 
conduction electron t-matrix is thus expressed as  
\begin{equation}
 t_{k,-k'}(\omega_n)=\frac{V^2}{N} G_{\rm vec}(\omega_n).
\end{equation}
It is easily seen that this t-matrix {\it only} depends on the
Green function for the impurity vector field $G_{\rm vec}(\tau)$. The 
implication of this relation is that the conduction electron 
resistivity will be determined by the $G_{\rm vec}(\tau)$ only.
Moreover, there is also a general relation between the impurity vector 
propagator and the cross correlation function,
\begin{equation}
 G_{k,d}(\omega_n)\equiv
\langle\langle\Psi_{\alpha}(k)|d_{\alpha}\rangle\rangle
 =\frac{iV}{\sqrt{N}}\frac{G_{\rm vec}(\omega_n)}
          {(i\omega_n-\epsilon_k)} \hspace{1cm} \alpha=1,2,3.
\end{equation} 
This relation becomes useful when we calculate various two-particle correlation
functions for the conduction electrons, because we can relate the
conduction electron correlation functions at the impurity site to the impurity
correlation functions.

In addition, we note  that under the simple transformation $S=\sqrt{2}d_0$, 
the model Hamiltonian  transforms according to  
\begin{equation}
     S H(U) S^{-1} = H(-U).
\end{equation}
Since the sign change of the on-site Hubbard interaction corresponds to 
the exchange of the impurity charge and spin degrees of freedom, or the
exchange of the impurity spin and `isospin' density operators,  
the particle-hole symmetry is kept in the present model, and 
the the chemical potential is pinned at the Fermi level.

\section{The unperturbed Hamiltonian}

Before considering the effects of interactions, it is very constructive to 
examine the unperturbed model Hamiltonian.
\begin{equation}
H_0= it\sum_n\sum_{\alpha=0}^3\Psi_{\alpha}(n+1)\Psi_{\alpha}(n)
        +iV\sum_{\alpha=1}^3\Psi_{\alpha}(0)d_{\alpha}.
\end{equation}
This is an exactly soluble Hamiltonian in which only the vector field of the 
conduction electrons hybridizes with the impurity vector field, while the 
scalar field is free. The impurity Green functions are defined as follows:
\begin{equation}
 G_0(\tau)=-\langle T_{\tau}d_0(\tau)d_0(0)\rangle, \hspace{1cm}
 G_{\alpha}(\tau)=-\langle T_{\tau}d_{\alpha}(\tau)d_{\alpha}(0)\rangle.
\end{equation}
It is straight forward to calculate their Fourier counterparts,
\begin{equation}
G_0(\omega_n)=\frac{1}{i\omega_n}, \hspace{1cm}
G_{\alpha}(\omega_n)=\frac{1}{i\omega_n+i\Delta{\rm sgn}\omega_n},
\end{equation}
where $\Delta=\pi\rho V^2$ is the hybridization width, $\rho=(hv_f)^{-1}$ 
is the conduction electron density of states, and $\omega_n=(2n+1)\pi/\beta$ 
is the fermionic Matsubara frequency. Here we find that the impurity scalar
propagator is a fermionic zero mode with $G_0(\tau)=-{\rm sgn}\tau/2$. 
Moreover, both impurity scalar and vector field 
propagators are odd in their arguments. 

The local impurity spectral function can be evaluated as 
\begin{equation}
  A_d(\omega)=\frac{3}{2\pi}\frac{\Delta}{\omega^2+\Delta^2}
               +\frac{1}{2}\delta(\omega),
\end{equation}
which reveals the basic physics of the unperturbed Hamiltonian that half 
of the impurity degree of freedom is free, while the rest of the degrees of 
freedom couple to the conduction electrons. A similar impurity spectral 
function was also found in the single-impurity two-channel Kondo model 
\cite{ek}. The change of 
free energy due to the hybridization with the impurity can be calculated as
\begin{equation}
  F^{(0)}_{\rm imp}=\frac{3}{2\pi}\int^{\infty}_{-\infty}f(\omega)
   {\tan}^{-1}\left (\frac{\Delta}{\omega}\right )d\omega -\frac{T}{2}\ln2,
\end{equation}
where $f(\omega)$ is Fermi distribution function, and the impurity residual 
entropy is found to be $\ln \sqrt{2}$, revealing that the ground state 
of $H_0$ is a two-fold degenerate state, which is a consequence of the 
spin-isospin symmetry with $[H_0, S]=0$ for  $S=\sqrt{2}d_0$.

In addition, we can calculate the spin and charge density-density 
correlation functions to find out what kind of low-energy behavior is described
by $H_0$. In the Majorana representation, the impurity spin and charge density
operators are expressed as
$$ S_d^z = \frac{1}{2} ( d^{\dag}_{\uparrow}d_{\uparrow} 
  - d^{\dag}_{\downarrow}d_{\downarrow})
   =-\frac{i}{2}(d_1d_2 - d_0d_3),$$
$$ n_d=\frac{1}{2} (d^{\dag}_{\uparrow}d_{\uparrow}
        +d^{\dag}_{\downarrow}d_{\downarrow}-1)
      =-\frac{i}{2}(d_1d_2 + d_0d_3).$$
The spin and charge density-density correlation functions are equal to 
\begin{equation}
 \langle T_{\tau}S^z_d(\tau)S_d^z(0)\rangle 
 =\langle T_{\tau}n_d(\tau)n_d(0)\rangle
=\frac{1}{4}\left [G_{\alpha}^2(\tau)+G_{\alpha}(\tau)G_0(\tau)\right ].
\end{equation}
Their corresponding Fourier transform is 
\begin{equation}
 \chi^{\rm imp}_{\rho,\sigma}(\omega_n)
  =\frac{1}{4\beta}\sum_{\omega_{n'}}
   \left [ G_{\alpha}(\omega_{n'})G_{\alpha}(\omega_n-\omega_{n'})
          + G_{\alpha}(\omega_{n'})G_0(\omega_n - \omega_{n'})\right ],
\end{equation}
where $\omega_n=2n\pi/\beta$ is the boson-type Matsubara frequency.  
The first term in the brackets corresponds to the normal FL-like     
density-density correlation forms, but the second term is singular. 
As far as the singularity is concerned, the imaginary part of the
spectral functions is obtained by completing the summation over frequency:
$$ {\rm Im}\chi^{\rm imp} _{\rho,\sigma}(\omega,T)
=-\frac{1}{8}\frac{\Delta}{\omega^2+\Delta^2}\tanh\left(\frac{\omega}{2T}\right)
\sim \left \{  
\begin{array}{ll}
 \frac{\omega}{T}, & \omega\ll T \\
 {\rm const.}, & \Delta\gg\omega\gg T
\end{array} 
 \right . $$
This kind of behavior was assumed in the marginal FL phenomenology \cite{varma}
to describe the normal state of the high temperature superconductors,
where the standard FL theory does not appear to explain the experimental
observations. The 
conduction electron charge and spin density-density correlation functions at 
the impurity site can also be calculated, but their spectral functions
do not display any kind of singular behavior.

It is clear  that the  Hamiltonian with $U=0$ has singular behavior
due to the degeneracy of the  ground state and 
 does not describe a non-interacting Fermi liquid.  In the
perturbation theory we develop in the next section, we make an
expansion about this unusual weak-coupling limit.

\section{Perturbation formalism of the model Hamiltonian}

\subsection{Partition function and free energy}

Now we  consider the full Hamiltonian for the compactified Anderson 
single-impurity model, 
\begin{eqnarray}
 && H=H_0+H_I, \nonumber \\
 && H_0=it\sum_n\sum_{\alpha=0}^3\Psi_{\alpha}(n+1)\Psi_{\alpha}(n)
        +iV\sum_{\alpha=1}^3\Psi_{\alpha}(0)d_{\alpha},
  \nonumber \\
 && H_I = -Ud_0d_1d_2d_3.
\end{eqnarray}
The partition function for this Hamiltonian can be expanded as follows:
\begin{eqnarray}
 && Z= Z_0 \sum_{n=0}^{\infty} (-1)^n \int_0^{\beta} d\tau_{n} 
  \int_0^{\tau_{n}} d\tau_{n-1}....\int_0^{\tau_{2}} d\tau_{1} 
   \langle H_I(\tau_{n})H_I(\tau_{n-1})...H_I(\tau_{1})\rangle
 \nonumber \\ && \hspace{0.5cm}
  =Z_0 \sum_{n=0}^{\infty} {U^n} \int_0^{\beta} d\tau_n
 \int_0^{\tau_n} d\tau_{n-1}....\int_0^{\tau_2} d\tau_1
     F_n(\tau_n,\tau_{n-1},...,\tau_1),
\end{eqnarray}
where $Z_0$ denotes the partition function for the unperturbed Hamiltonian, and
$\langle...\rangle$ means the thermodynamic average is carried out over the 
unperturbed 
part $H_0$. Noting that in $H_0$ the four Majorana fermion components of the 
local impurity decouple completely, 
$F_n(\tau_n,\tau_{n-1},...,\tau_1)$ can be factorized as
\begin{equation}
 \langle d_0(\tau_n)d_0(\tau_{n-1})...d_0(\tau_1)\rangle \prod_{\alpha=1}^3
  \langle d_{\alpha}(\tau_n)d_{\alpha}(\tau_{n-1})...d_{\alpha}(\tau_1)\rangle.
\end{equation}
Here both the impurity scalar and vector field single-particle Green function 
$G_0(\tau)$ and $G_{\alpha}(\tau)$ ($\alpha=1,2,3$) are free propagators. 
When the Wick theorem is implemented, it can be verified order by order,
that each Majorana fermion expectation average can be represented by a Pfaffian
determinant \cite{gh}. For the expectation values of the impurity vector 
operators, the 
Pfaffian determinant is defined by the square root of an antisymmetric 
determinant composed of the impurity vector propagator $G_{\alpha}(\tau)$,
  $$\left |
  \begin{array}{lllll}
  0, & G_{\alpha}(\tau_1-\tau_2),& G_{\alpha}(\tau_1-\tau_3),& ......,&
   G_{\alpha}(\tau_1-\tau_n) \\
  G_{\alpha}(\tau_2-\tau_1),& 0,& G_{\alpha}(\tau_2-\tau_3),&......,&
   G_{\alpha}(\tau_2-\tau_n) \\
 \vdots & \vdots & \vdots &  & \vdots \\
  G_{\alpha}(\tau_n-\tau_1),& G_{\alpha}(\tau_n-\tau_2),&
  G_{\alpha}(\tau_n-\tau_3),& .......,& 0
  \end{array}
  \right |= \backslash D_n(\tau_1,\tau_2,....,\tau_n)|^2.$$ 
As shown in Section II, the impurity vector field propagators are odd in their 
arguments,
$G_{\alpha}(\tau_r-\tau_s)+G_{\alpha}(\tau_s-\tau_r)=0$ and
$G_{\alpha}(\tau_r=\tau_s)=0$.  
The Pfaffian determinant $\backslash D_n(\tau_1,\tau_2,..,\tau_n)|$ is given by 
  $$ \left .  
  \begin{array}{llll}
   |\hspace{.1cm} 
   G_{\alpha}(\tau_1-\tau_2),& G_{\alpha}(\tau_1-\tau_3),& ....,&
   G_{\alpha}(\tau_1-\tau_n) \\
   & G_{\alpha}(\tau_2-\tau_3),&....,& G_{\alpha}(\tau_2-\tau_n) \\ 
   &                           &    & \vdots         \\
   &                           &    & G_{\alpha}(\tau_{n-1}-\tau_n)
  \end{array} \right | 
 = \sum \pm G_{\alpha}(\tau_1-\tau_a)G_{\alpha}(\tau_b-\tau_c)......
                      G_{\alpha}(\tau_l-\tau_m),$$
where the subscripts $1,a,b,c,......,l,m$ of each term under the summation are
a permutation of the first $n$ integers, each Green function 
$G_{\alpha}(\tau_r-\tau_s)$ has $s>r$, all different terms of this type are 
included, and the total number of terms is 
$(n-1)(n-3)\cdot\cdot\cdot 5\cdot 3\cdot 1$. The sign
attached to each term is positive or negative depending on whether the 
permutation
is even or odd. In addition, the basic property of Pfaffian determinant
is that all odd-order determinants identically vanish, and all the expansion 
terms we should consider are even-order determinants. For instance, the fourth
Pfaffian determinant $\backslash D_4(\tau_1,\tau_2,\tau_3,\tau_4)|$ is given by
 $$ \left .
  \begin{array}{lll}
  |\hspace{0.1cm}
 G_{\alpha}(\tau_1,\tau_2),& G_{\alpha}(\tau_1,\tau_3),& 
      G_{\alpha}(\tau_1,\tau_4)\\
                  & G_{\alpha}(\tau_2,\tau_3),& G_{\alpha}(\tau_2,\tau_4)\\
                  &                 & G_{\alpha}(\tau_3,\tau_4)
  \end{array} \right | $$
  $$ = G_{\alpha}(\tau_1,\tau_2)G_{\alpha}(\tau_3,\tau_4)- 
G_{\alpha}(\tau_1,\tau_3)G_{\alpha}(\tau_2,\tau_4)
           + G_{\alpha}(\tau_1,\tau_4)G_{\alpha}(\tau_2,\tau_3).$$

On the other hand, the impurity scalar field propagator has a special form, 
 $G_0(\tau)=-{\rm sgn}\tau/2$, its corresponding expectation is 
trivial to calculate because the imaginary time sequence has been assumed
$\beta>\tau_{2n}>\tau_{2n-1}>\cdot\cdot\cdot\cdot>\tau_2>\tau_1>0$. 
\begin{equation}
 \langle d_0(\tau_{2n})d_0(\tau_{2n-1})...d_0(\tau_1)\rangle 
 =\left(\frac{1}{2}\right)^n.
\end{equation}
Therefore, the partition function for the compactified  Anderson single-impurity
model $H$ can be expressed as the cube of the Pfaffian determinant with 
elements corresponding to the
impurity vector field single-particle Green function at different times:
\begin{equation}
 Z/Z_0= \sum_{n=0}^{\infty} \left(\frac{U}{\sqrt{2}}\right)^{2n} 
  \int_0^{\beta} d\tau_{2n}
 \int_0^{\tau_{2n}} d\tau_{2n-1}....\int_0^{\tau_2} d\tau_1
 \left\{ \backslash D_{2n}(\tau_1,\tau_2,...,\tau_{2n})|\right \}^3.
\end{equation}
For the ordinary symmetric Anderson model with $V_0=V$, the
power of the Pfaffian determinant in the partition function is four rather than 
three, and in fact there is no need to introduce Pfaffian determinant 
in that case \cite{yy,yamada}. Thus one power of the Pfaffian determinant in the
partition function corresponds to each of the Majorana fermions involved in 
the hybridization.
Using the linked cluster theorem, we find that the free energy change due to 
the local impurity can be expressed in power series in the parameter $U$ 
\begin{equation}\label{F}
 F_{\rm imp}=F^{(0)}_{\rm imp} 
 -\sum_{n=1}^{\infty}\left(\frac{U}{\sqrt{2}}\right)^{2n}
\frac{1}{\beta}
 \int_0^{\beta}d\tau_{2n}\int_0^{\tau_{2n}}d\tau_{2n-1}
  ....\int_0^{\tau_2}d\tau_1
\left\{ \backslash D_{2n}(\tau_1,\tau_2,...,\tau_{2n})|\right\} ^3_{l},
\end{equation} 
where the subscript $l$ on the bracket in the above equation indicates that 
only linked diagrams are to be considered. The extra impurity specific heat 
can be calculated systematically from this free energy expression.

\subsection{Single-particle Green functions and self energies}

Now we  consider the perturbed impurity single-particle correlation 
functions. For the impurity scalar field $d_0$, the perturbed Green function is 
defined as
\begin{equation}
  G_{\rm sc}(\tau,\tau')=-\frac{\langle T_{\tau}d_0(\tau)d_0(\tau')
     exp\left[-\int_0^{\beta}H_I(\tau)d\tau\right]\rangle }
     {\langle T_{\tau}exp\left[-\int_0^{\beta}H_I(\tau)d\tau\right]\rangle },
\end{equation}
and its Fourier transform is given by
\begin{equation}
 G_{\rm sc}(\omega_n)=\frac{1}{\beta}\int_0^{\beta}d\tau \int_0^{\beta}d\tau' 
               G_{\rm sc}(\tau,\tau')e^{i\omega_n(\tau-\tau')}.
\end{equation}
In a similar way to our early derivation of the partition function, the 
perturbed scalar field propagator can be expanded in powers of $U$:
\begin{eqnarray}
 && G_{\rm sc}(\omega_n)=G_0(\omega_n)+ 
 \sum_{n=1}^{\infty}\frac{U^{2n}}{\beta}
 \int_0^{\beta}d\tau\int_0^{\beta}d\tau' 
  \int_0^{\beta}d\tau_{2n}\int_0^{\tau_{2n}}d\tau_{2n-1}...
   \int_0^{\tau_2}d\tau_1 e^{i\omega_n(\tau-\tau')}
\nonumber \\ && \hspace{2cm}
 \left\{\langle T_{\tau}d_0(\tau)d_0(\tau')d_0(\tau_{2n})....d_0(\tau_1)\rangle
      \backslash D_{2n}(\tau_1,\tau_2,...,\tau_{2n})|^3\right\}_{l}.
\end{eqnarray}
For the special form of $G_0(\tau)=-{\rm sgn}\tau/2$, the expectation value
of the impurity scalar field propagator can be calculated as
\begin{eqnarray}
&& -\langle T_{\tau}d_0(\tau)d_0(\tau')d_0(\tau_{2n})....d_0(\tau_1)\rangle
 \nonumber \\ && 
=\left(\frac{1}{2}\right)^{n-1}
 \sum_{i<j}(-1)^{i+j} [ G_0(\tau-\tau_i)G_0(\tau'-\tau_j)
        -G_0(\tau-\tau_j)G_0(\tau'-\tau_i)]
   +\left(\frac{1}{2}\right)^n G_0(\tau-\tau').
\end{eqnarray}
Then completing the integrals over $\tau$ and $\tau'$, 
we obtain the scalar field Green function in terms of the following equation,  
\begin{equation}
 G_{\rm sc}(\omega_n)
=G_0(\omega_n) + G_0(\omega_n)\Sigma'_{\rm sc}(\omega_n)G_0(\omega_n),
\end{equation}
where an improper self-energy is represented as 
\begin{eqnarray}
&& \Sigma'_{\rm sc}(\omega_n)= 
\sum_{n=1}^{\infty}\frac{4i}{\beta}\left(\frac{U}{\sqrt{2}}\right)^{2n}
 \int_0^{\beta}d\tau_{2n}\int_0^{\tau_{2n}}d\tau_{2n-1}
  ....\int_0^{\tau_2} d\tau_1
\nonumber \\ && \hspace{1cm}
 \left \{ \sum_{i<j}(-1)^{i+j}[\sin\omega_n(\tau_i-\tau_j)]
\left\{ \backslash D_{2n}(\tau_1,\tau_2,...,\tau_{2n})|\right\}^3 \right \}_{l}.
\end{eqnarray}

For the impurity vector field $d_{\alpha}$ with ($\alpha=1,2,3$), the 
definition of its Green function $G_{\rm vec}(\tau,\tau')$ is similar to the 
scalar field, and its Fourier counterpart $G_{\rm vec}(\omega_n)$ can 
also be expanded in a power series in $U$:
\begin{eqnarray}
&& G_{\rm vec}(\omega_n)=G_{\alpha}(\omega_n)+
 \sum_{n=1}^{\infty}\left(\frac{U}{\sqrt{2}}\right)^{2n}\frac{1}{\beta}
 \int_0^{\beta}d\tau\int_0^{\beta}d\tau' 
 \int_0^{\beta}d\tau_{2n}....\int_0^{\tau_2}d\tau_1
 e^{i\omega_n(\tau-\tau')}
\nonumber \\ && \hspace{2cm}
 \left \{\langle T_{\tau}d_{\alpha}(\tau)d_{\alpha}(\tau')
 d_{\alpha}(\tau_{2n})....d_{\alpha}(\tau_1)\rangle
      \backslash D_{2n}(\tau_1,\tau_2,...,\tau_{2n})|^2 \right\}_{l}.
\end{eqnarray}
The remaining  expectation average over $d_{\alpha}$ can be calculated as 
\begin{eqnarray}
 && -\langle T_{\tau}d_{\alpha}(\tau)d_{\alpha}(\tau')d_{\alpha}(\tau_{2n})....
          d_{\alpha}(\tau_1)\rangle
 \nonumber \\ &&
=\sum_{i<j}(-1)^{i+j}[G_{\alpha}(\tau-\tau_i)G_{\alpha}(\tau'-\tau_j)
    -G_{\alpha}(\tau-\tau_j)G_{\alpha}(\tau'-\tau_i)] 
     \backslash D_{2n}^{ij}|+G_{\alpha}(\tau-\tau')\backslash D_{2n}|, 
\end{eqnarray}
where $\backslash D_{2n}^{ij}|$ is the so-called {\it cofactor} of Pfaffian 
determinant $\backslash D_{2n}|$, which  is  a Pfaffian determinant
of the same type but with  $\tau_i$ and $\tau_j$ removed from the time 
sequence. 
Then we can perform the integrals over $\tau$ and $\tau'$ and get the 
impurity vector field Green function,
\begin{equation}
 G_{\rm vec}(\omega_n)=G_{\alpha}(\omega_n)
 +G_{\alpha}(\omega_n)\Sigma'_{\rm vec}(\omega_n)G_{\alpha}(\omega_n),
\end{equation}
from which the corresponding improper self-energy is extracted as 
\begin{eqnarray}
&& \Sigma'_{\rm vec}(\omega_n) =
\sum_{n=1}^{\infty}\frac{2i}{\beta}\left(\frac{U}{\sqrt{2}}\right)^{2n}
 \int_0^{\beta}d\tau_{2n}\int_0^{\tau_{2n}}d\tau_{2n-1}
  ....\int_0^{\tau_2} d\tau_1
 \nonumber \\ && \hspace{1cm}
\left \{ \sum_{i<j}(-1)^{i+j}[\sin\omega_n(\tau_i-\tau_j)]
       \backslash D_{2n}^{ij}(\tau_1,\tau_2,...,\tau_{2n})|
\left\{ \backslash D_{2n}(\tau_1,\tau_2,...,\tau_{2n})|\right\}^2 \right \}_{l}.
\end{eqnarray}
 
The self-energies usually used are the
proper self energies, which are related to the improper ones by 
\begin{eqnarray}
&& \Sigma_{\rm sc}(\omega_n)
=\Sigma'_{\rm sc}(\omega_n)[1+G_0(\omega_n)\Sigma'_{\rm sc}(\omega_n)]^{-1};
\nonumber \\ &&
\Sigma_{\rm vec}(\omega_n)
=\Sigma'_{\rm vec}(\omega_n)
  [1+G_{\alpha}(\omega_n)\Sigma'_{\rm vec}(\omega_n)]^{-1}.
\end{eqnarray}
As far as the leading-order perturbative contributions are concerned, it is 
not necessary to 
distinguish the improper self energies from the proper ones. 

\subsection{Two-particle Green functions and vertex functions}

The method of evaluating the single-particle correlation functions 
can be applied further to the impurity two-particle Green functions. Then the 
two-particle vertex functions can be calculated to  give  a more complete
picture of the physical behavior of this  model. 
Generally, three different two-particle correlation functions can be defined.
The first one is given by 
\begin{eqnarray}
&&  G^{II}_{0,1,2,3}(\omega,\omega',\omega'',\omega''')
 \nonumber \\ &&  
=\int_{0}^{\beta}...\int_{0}^{\beta}d\tau d\tau' d\tau'' d\tau'''
  \langle T_{\tau}d_0(\tau)d_1(\tau')d_2(\tau'')d_3(\tau''')\rangle
   e^{i(\omega\tau+\omega'\tau'+\omega''\tau''+\omega'''\tau''')}.
\end{eqnarray}
Using  the same strategies we  used earlier in calculating the impurity 
single-particle Green functions, we find that
\begin{eqnarray}
&& G^{II}_{0,1,2,3}(\omega,\omega',\omega'',\omega''')
\nonumber \\ && 
=- \sqrt{2}G_0(\omega)G_{\alpha}(\omega')G_{\alpha}(\omega'')
   G_{\alpha}(\omega''')
  \sum_{n=0}^{\infty}\left(\frac{U}{\sqrt{2}}\right)^{2n+1}
\int_0^{\beta}d\tau_{2n+1}\int_0^{\tau_{2n+1}}d\tau_{2n}
  ....\int_0^{\tau_2} d\tau_1
\nonumber \\ && \hspace{.5cm}
 \left \{ \sum_{i,j}\sum_{i',j'}(-1)^{i+j+i'+j'}
      e^{i(\omega\tau_i+\omega'\tau_j+\omega''\tau_{i'}+\omega'''\tau_{j'})}
  \backslash D_{2n+1}^j|\backslash D_{2n+1}^{i'}|\backslash D_{2n+1}^{j'}|
   \right \}_l.
\end{eqnarray}
 The corresponding improper vertex function can be defined as 
\begin{eqnarray}
&& \Gamma'_{0,1,2,3}(\omega,\omega',\omega'',\omega''')=-
 \sum_{n=0}^{\infty}\frac{\sqrt{2}}{\beta}\left(\frac{U}{\sqrt{2}}\right)^{2n+1}
 \int_0^{\beta}d\tau_{2n+1}\int_0^{\tau_{2n+1}}d\tau_{2n}
  ....\int_0^{\tau_2}d\tau_1
\nonumber \\ && \hspace{.5cm}
 \left \{ \sum_{i,j}\sum_{i',j'}(-1)^{i+j+i'+j'}
      e^{i(\omega\tau_i+\omega'\tau_j+\omega''\tau_{i'}+\omega'''\tau_{j'})}
  \backslash D_{2n+1}^j|\backslash D_{2n+1}^{i'}|\backslash D_{2n+1}^{j'}|
   \right \}_l.
\end{eqnarray}
When all the frequencies are set to zero, this expression simplifies 
to 
\begin{eqnarray}
 && \Gamma'_{0,1,2,3}(0,0,0,0)
\nonumber \\ &&
=\sum_{n=0}^{\infty}\frac{\sqrt{2}}{\beta}\left(\frac{U}{\sqrt{2}}\right)^{2n+1}
 \int_0^{\beta}d\tau_{2n+1}\int_0^{\tau_{2n+1}}d\tau_{2n}
  ....\int_0^{\tau_2} d\tau_1
 \left \{ \sum_i(-1)^i
       \backslash D_{2n+1}^i(\tau_1,...,\tau_{2n+1})|\right\}^3_l.
\end{eqnarray}
This vertex function will play an important role 
when we  apply the multiplicative renormalization group to
our perturbation expansion in section VII.

The above treatment can be applied to the following two two-particle 
Green functions straight forwardly:
$$ 
\langle T_{\tau}d_0(\tau)d_0(\tau')d_{\alpha}(\tau'')d_{\alpha}(\tau''')\rangle
 \hspace{.5cm} {\rm and} \hspace{.5cm}
 \langle T_{\tau}d_{\alpha}(\tau)d_{\alpha}(\tau')d_{\beta}(\tau'')
        d_{\beta}(\tau''')\rangle,$$
here $\alpha, \beta = 1,2,3$, and $\alpha\neq \beta$.
Their resulting equations of motion turn out to be 
\begin{eqnarray}
&& G^{II}_{0,0,\alpha,\alpha}(\omega,\omega',\omega'',\omega''')
 \nonumber \\ &&
  =\beta^2 G_0(\omega)G_{\alpha}(\omega'')
  \delta_{\omega',-\omega}\delta_{\omega''',-\omega''}
 +\beta G_0(\omega)G_0(\omega')G_{\alpha}(\omega'')G_{\alpha}(\omega''')
 \Gamma'_{0,0,\alpha,\alpha}(\omega,\omega',\omega'',\omega''');
 \nonumber \\ &&
G^{II}_{\alpha,\alpha,\beta,\beta}(\omega,\omega',\omega'',\omega''')
 \nonumber \\ &&
 =\beta^2 G_{\alpha}(\omega)G_{\alpha}(\omega'')
    \delta_{\omega',-\omega}\delta_{\omega''',-\omega''}
 +\beta G_{\alpha}(\omega)G_{\alpha}(\omega')G_{\alpha}(\omega'')
  G_{\alpha}(\omega''')
 \Gamma'_{\alpha,\alpha,\beta,\beta}(\omega,\omega',\omega'',\omega'''),
\end{eqnarray} 
where the improper vertex functions are given by 
\begin{eqnarray}
&& \Gamma'_{0,0,\alpha,\alpha} (\omega,\omega',\omega'',\omega''')
 = \sum_{n=0}^{\infty}\left(\frac{U}{\sqrt{2}}\right)^{2n}
 \frac{2}{\beta}
 \int_0^{\beta} d\tau_{2n}\int_0^{\tau_{2n}}d\tau_{2n-1}
  ....\int_0^{\tau_2} d\tau_{1}
  \left \{ \sum_{i<j}\sum_{i'<j'} (-1)^{i+j+i'+j'}\right.
 \nonumber \\ && \hspace{.5cm}
\left. \left[e^{i(\omega\tau_i+\omega'\tau_j)}
          -e^{i(\omega\tau_j+\omega'\tau_i)}\right]
  \left[ e^{i(\omega''\tau_{i'}+\omega'''\tau_{j'})} -
     e^{i(\omega''\tau_{j'}+\omega'''\tau_{i'} ) } \right]
  \backslash D_{2n}^{i'j'}| \backslash D_{2n}|^2 \right\}_{l};
\end{eqnarray}
\begin{eqnarray}
&& \Gamma'_{\alpha,\alpha,\beta,\beta}(\omega,\omega',\omega'',\omega''')
 = \sum_{n=0}^{\infty}\left(\frac{U}{\sqrt{2}}\right)^{2n}
 \frac{1}{\beta}
 \int_0^{\beta} d\tau_{2n}\int_0^{\tau_{2n}}d\tau_{2n-1}
  ....\int_0^{\tau_2} d\tau_{1}
 \left \{ \sum_{i<j}\sum_{i'<j'}(-1)^{i+j+i'+j'}\right. 
  \nonumber \\ && \hspace{.5cm}
\left. \left[e^{i(\omega\tau_i+\omega'\tau_j)}
          -e^{i(\omega\tau_j+\omega'\tau_i)}\right]
  \left[ e^{i(\omega''\tau_{i'} + \omega'''\tau_{j'} ) } -
     e^{i(\omega''\tau_{j'} + \omega'''\tau_{i'} ) } \right]
   \backslash D_{2n}^{ij}| \backslash D_{2n}^{i'j'}|
        \backslash D_{2n}| \right\}_{l}.
\end{eqnarray}
These two improper vertex functions are antisymmetric
in their arguments and
$$ \Gamma'_{0,0,\alpha,\alpha}(0,0,\omega'',\omega''')=
   \Gamma'_{0,0,\alpha,\alpha}(\omega,\omega',0,0)=
  \Gamma'_{\alpha,\alpha,\beta,\beta}(0,0,\omega'',\omega''')=
  \Gamma'_{\alpha,\alpha,\beta,\beta}(\omega,\omega',0,0)=0.$$

Since the last two vertex functions vanish in the zero frequency 
limit, we expect only $\Gamma_{0,1,2,3}$ to be important in 
determining the two-particle interactions in the low-energy regime. 
It is straightforward to relate the improper and proper vertex function,
for example,
\begin{equation}
 \Gamma_{0,1,2,3}(\omega,\omega',\omega'',\omega''')=
 \frac{1}{z_{\rm sc}}\frac{1}{z^3_{\rm vec}}
 \Gamma'_{0,1,2,3}(\omega,\omega',\omega'',\omega'''),
\end{equation}
where $z_{\rm sc}(\omega_n)=G_{\rm sc}(\omega_n)/G_0(\omega_n)$ and
$z_{\rm vec}(\omega_n)=G_{\rm vec}(\omega_n)/G_{\alpha}(\omega_n)$ denote the 
wave function renormalization factors for the impurity scalar and vector 
components, respectively.  

\subsection{Dyson equations of the impurity self-energies}

The impurity scalar and vector self energies $\Sigma_{\rm sc}$ and
$\Sigma_{\rm vec}$ have already been derived in terms of the Pfaffian
determinants and their co-factors, which are functions of
the unperturbed propagators. However,  when it is necessary to sum an
infinite series, instead of just calculating the first few diagrams, it is
usually more convenient to express the impurity self energies in terms of
the vertex function $\Gamma_{0,1,2,3}$. On the other hand, these two impurity
self energies are also dependent of the vertex function $\Gamma_{0,1,2,3}$.
The procedures are illustrated by considering $\Sigma_{\rm sc}$ as an example.

Since $\Sigma_{\rm sc}$ is even function of $U$, there are no first-order
perturbation corrections. Also there are no higher-order diagrams in which the
self energy is connected to the basic propagator $G_{\rm sc}$ by a single
first-order vertex function 
$\Gamma^{(1)}_{0,1,2,3}(\omega,\omega',\omega'',\omega''')
=-U\delta_{\omega+\omega'+\omega''+\omega''',0}$.
The second-order perturbative corrections to the self energy
$\Sigma_{\rm sc}$ will be derived in Section VII, here we just use the result,
\begin{equation}
 \Sigma^{(2)}_{\rm sc}(\omega_n)
 =-\frac{U^2}{\beta^2}\sum_{\omega_1,\omega_2}G_{\alpha}(\omega_1)
            G_{\alpha}(\omega_2)G_{\alpha}(\omega_n-\omega_1-\omega_2).
\end{equation}
It is not hard to see that all the higher order  diagrams for $\Sigma_{\rm sc}$
can be
obtained from this second-order diagram by insertion of self energy parts into
internal unperturbed propagators $G_{\alpha}$ and replacement of the square on
the right by the proper vertex
function $\Gamma_{0,1,2,3}$. This diagram is displayed in Fig.1a. Then the
Dyson equation of the impurity scalar self energy is derived in terms of
the impurity vertex function,
\begin{eqnarray}\label{sc}
&& \Sigma_{\rm sc}(\omega_n)
\nonumber \\ &&
 =\frac{U}{\beta^2}\sum_{\omega_1,\omega_2} G_{\rm vec}(\omega_1)
    G_{\rm vec}(\omega_2)G_{\rm vec}(\omega_n-\omega_1-\omega_2)
      \Gamma_{0,1,2,3}(\omega_2,\omega_n-\omega_1-\omega_2;\omega_1,-\omega_n).
\end{eqnarray}

In a similar way, we can build up the Dyson equation for the impurity vector
self energy $\Sigma_{\rm vec}$ from its second order perturbative
corrections, which will also be derived in section VII,
\begin{equation}
 \Sigma^{(2)}_{\rm vec}(\omega_n)
 =-\frac{U^2}{\beta^2}\sum_{\omega_1,\omega_2}G_{\alpha}(\omega_1)
            G_{\alpha}(\omega_2)G_0(\omega_n-\omega_1-\omega_2).
\end{equation}
The corresponding Dyson equation is given by
\begin{eqnarray}\label{vec}
 && \Sigma_{\rm vec}(\omega_n)
\nonumber \\ &&
 =\frac{U}{\beta^2}\sum_{\omega_1,\omega_2} G_{\rm vec}(\omega_1)
    G_{\rm vec}(\omega_2)G_{\rm sc}(\omega_n-\omega_1-\omega_2)
      \Gamma_{0,1,2,3}(\omega_2,\omega_n-\omega_1-\omega_2;\omega_1,-\omega_n),
\end{eqnarray}
the corresponding diagram is described by Fig.1b.
Eq.(\ref{sc}) and Eq.(\ref{vec}) are two basic equations of the
model Hamiltonian and also exhibit the relations between the impurity self
energies and the two-particle vertex function which is finite in the zero
frequency limit.

\section{Dynamical susceptibilities}
 
\subsection{Impurity spin and charge dynamical susceptibilities}

As noted earlier the impurity charge and spin density operators are 
defined by
\begin{equation}
 S_d^z =-\frac{i}{2}(d_1d_2 - d_0d_3),\hspace{.5cm}
 n_d=-\frac{i}{2}(d_1d_2 + d_0d_3).
\end{equation}
Using these expression the 
 impurity spin and charge density-density correlation functions in the
Fourier space can be  expressed in the form,
\begin{eqnarray}
 && \chi_{\sigma}^{\rm imp}(\omega_n)=\left(\frac{g\mu}{2}\right)^2
  \{\tilde{\chi}_{\rm even}(\omega_n)+\tilde{\chi}_{\rm odd}(\omega_n)\},
 \nonumber \\ &&
   \chi_{\rho}^{\rm imp}(\omega_n)=\frac{1}{4}
     \{\tilde{\chi}_{\rm even}(\omega_n)-\tilde{\chi}_{\rm odd}(\omega_n)\},
\end{eqnarray}
with   
\begin{eqnarray}
&& \tilde{\chi}_{\rm even}(\omega_n) 
 =\frac{1}{\beta}\int_0^{\beta}d\tau\int_0^{\beta}d\tau'
   \{ \langle T_{\tau}d_1(\tau)d_1(\tau')d_2(\tau)d_2(\tau')\rangle
    + \langle T_{\tau}d_0(\tau)d_0(\tau')d_3(\tau)d_3(\tau')\rangle\},
\nonumber \\
&& \tilde{\chi}_{\rm odd}(\omega_n)
 =\frac{2}{\beta}\int_0^{\beta}d\tau\int_0^{\beta}d\tau'
     \langle T_{\tau}d_0(\tau)d_1(\tau')d_2(\tau')d_3(\tau)\rangle. 
\end{eqnarray}

Expressions for these correlation functions in terms of the Pfaffian 
determinants are rather complicated.  We can express them however in terms of 
the impurity improper self-energies and vertex functions, which is a more 
convenient form for the calculation of the low-order 
perturbation corrections. For the even part, the form is given by 
\begin{eqnarray}
&& \tilde{\chi}_{\rm even}(\omega_n)
\nonumber \\ &&
 =\frac{1}{\beta}\sum_{\omega_1}G_{\alpha}(\omega_1)
    G_{\alpha}(\omega_n-\omega_1)
+\frac{1}{\beta}\sum_{\omega_1}G_{\alpha}(\omega_1)G_0(\omega_n-\omega_1)
-\frac{2}{\beta}\sum_{\omega_1}G^{2}_{\alpha}(\omega_1)
    G_{\alpha}(\omega_n-\omega_1)\Sigma'_{\rm vec}(\omega_1)
\nonumber \\&& \hspace{.5cm}
 -\frac{1}{\beta}\sum_{\omega_1}G^{2}_{\alpha}(\omega_1)
    G_0(\omega_n-\omega_1)\Sigma'_{\rm vec}(\omega_1)
-\frac{1}{\beta}\sum_{\omega_1}G^{2}_0(\omega_1)
    G_{\alpha}(\omega_n-\omega_1)\Sigma'_{\rm sc}(\omega_1)
\nonumber \\ && \hspace{.5cm}
-\frac{1}{\beta^2}\sum_{\omega_1,\omega_2}G_{\alpha}(\omega_1)
 G_{\alpha}(\omega_2)G_{\alpha}(\omega_n-\omega_1)G_{\alpha}(\omega_n+\omega_2)
 \Gamma'_{\alpha,\alpha,\beta,\beta}(\omega_1,\omega_2;\omega_n-\omega_1,
 -\omega_n-\omega_2)
\nonumber \\ && \hspace{.5cm}
-\frac{1}{\beta^2}\sum_{\omega_1,\omega_2}G_0(\omega_1)
 G_0(\omega_2)G_{\alpha}(\omega_n-\omega_1)G_{\alpha}(\omega_n+\omega_2)
 \Gamma'_{0,0,\beta,\beta}(\omega_1,\omega_2;\omega_n-\omega_1,
 -\omega_n-\omega_2),
\end{eqnarray}
while the odd part is given by
\begin{eqnarray}
&& \tilde{\chi}_{\rm odd}(\omega_n)
\nonumber \\ &&
=\frac{2}{\beta^2}\sum_{\omega_1,\omega_2}G_0(\omega_1)
 G_{\alpha}(\omega_2)G_{\alpha}(\omega_n-\omega_1)G_{\alpha}(\omega_n+\omega_2)
 \Gamma'_{0,1,2,3}(\omega_1,\omega_2;\omega_n-\omega_1,-\omega_n-\omega_2),
\end{eqnarray}
where $G_0$ and $G_{\alpha}$ correspond to the unperturbed impurity scalar and 
vector propagators, respectively.

If the bosonic Matsubara frequency $\omega_n$ is taken to zero, the static
charge and spin susceptibilities are obtained. In fact, there is another way to
derive the impurity static susceptibilities. 
When an additional term $\delta H$ is added to the
 Hamiltonian corresponding to a coupling with an external
field, such as a coupling to  a uniform
magnetic field $\delta H_I=-i(g\mu_B h/2)(d_1d_2-d_0d_3)$
or a chemical potential term $\delta H=-i(\mu/2)(d_1d_2+d_0d_3)$,
 the partition function can be written in the form
\begin{equation}
 Z = {\rm Tr} \left \{ e^{-\beta H_0} T_{\tau}
    exp\left [ -\int_0^{\beta} H_I(\tau) d\tau \right ]
     exp\left [-\int_0^{\beta} \delta H_I(\tau') d\tau' \right ]
       \right \}.
\end{equation}
Then one can expand this partition function to second-order in $\delta H$ in a 
power series in $U$, and derive expressions for the static spin and charge 
susceptibilities:
$$ \chi_{\sigma}^{\rm imp}=\left(\frac{g\mu_B}{2}\right)^2\left \{
  \tilde{\chi}_{\rm even}+\tilde{\chi}_{\rm odd} \right\},
 \hspace{1cm} \chi_{\rho}^{\rm imp}
 =\frac{1}{4}\left \{\tilde{\chi}_{\rm even}-\tilde{\chi}_{\rm odd} \right\},$$
where $\tilde{\chi}_{\rm even}$ and $\tilde{\chi}_{\rm odd}$ are expressed in
terms of the Pfaffian determinant and its co-factors, respectively.
Since we do not use these forms to calculate the static susceptibilities, 
we shall not write them here.
 
It should  be pointed out that 
the above expressions for $\tilde{\chi}_{\rm even}(\omega_n)$ and 
$ \tilde{\chi}_{\rm odd}(\omega_n) $ can also be converted into alternative 
forms in terms of the perturbed propagators $G_{\rm sc}$ 
and $G_{\rm vec}$, and the proper vertex functions 
$\Gamma_{\alpha,\alpha,\beta,\beta}$ and $\Gamma_{0,0,\alpha,\alpha}$,
which are general expressions for the model Hamiltonian. 
\begin{eqnarray}
 && \tilde{\chi}_{\rm even}(\omega_n)
=\frac{1}{\beta}\sum_{\omega_1}G_{\alpha}(\omega_1)G_{\alpha}(\omega_n-\omega_1)
 +\frac{1}{\beta}\sum_{\omega_1}G_{\alpha}(\omega_1)G_0(\omega_n-\omega_1)
\nonumber \\&& \hspace{.5cm}
  -\frac{2}{\beta}\sum_{\omega_1}G_{\rm vec}(\omega_1)
       G_{\alpha}(\omega_n-\omega_1)
  -\frac{1}{\beta}\sum_{\omega_1}G_{\rm vec}(\omega_1)G_0(\omega_n-\omega_1)
  -\frac{1}{\beta}\sum_{\omega_1}G_{\rm sc}(\omega_1)
       G_{\alpha}(\omega_n-\omega_1)
\nonumber \\ && \hspace{.5cm}
  +\frac{2}{\beta}\sum_{\omega_1}G_{\rm vec}(\omega_1)
       G_{\rm vec}(\omega_n-\omega_1)
  +\frac{1}{\beta}\sum_{\omega_1}G_{\rm vec}(\omega_1)
       G_{\rm sc}(\omega_n-\omega_1)
  +\frac{1}{\beta}\sum_{\omega_1}G_{\rm sc}(\omega_1)
       G_{\rm vec}(\omega_n-\omega_1)
\nonumber \\ && \hspace{.5cm}
  -\frac{1}{\beta^2}\sum_{\omega_1,\omega_2}G_{\rm vec}(\omega_1)
 G_{\rm vec}(\omega_2)G_{\rm vec}(\omega_n-\omega_1)
     G_{\rm vec}(\omega_n+\omega_2)
 \Gamma_{\alpha,\alpha,\beta,\beta}(\omega_1,\omega_2;\omega_n-\omega_1,
 -\omega_n-\omega_2)
\nonumber \\ && \hspace{.5cm}
-\frac{1}{\beta^2}\sum_{\omega_1,\omega_2}G_{\rm sc}(\omega_1)
 G_{\rm sc}(\omega_2)G_{\rm vec}(\omega_n-\omega_1)
       G_{\rm vec}(\omega_n+\omega_2)
 \Gamma_{0,0,\beta,\beta}(\omega_1,\omega_2;\omega_n-\omega_1,
 -\omega_n-\omega_2);
\nonumber \\
&& \tilde{\chi}_{\rm odd}(\omega_n)
\nonumber \\ &&
=\frac{2}{\beta^2}\sum_{\omega_1,\omega_2}G_{\rm sc}(\omega_1)
 G_{\rm vec}(\omega_2)G_{\rm vec}(\omega_n-\omega_1)
       G_{\rm vec}(\omega_n+\omega_2)
 \Gamma_{0,1,2,3}(\omega_1,\omega_2;\omega_n-\omega_1,-\omega_n-\omega_2).
\end{eqnarray}
These two expressions can be described diagrammatically in Fig.2, where each 
term corresponds to one Feynman diagram.
 
By the way, the free energy (\ref{F}) can also be converted into a function
of the impurity perturbed propagators and vertex function  
$\tilde{\chi}_{\rm odd}(\omega_n)$ when we use the relation 
$$ \backslash D_{2n}(\tau_1,\tau_2,...,\tau_{2n})|
   = -\sum_i (-1)^{i} G_{\alpha}(\tau_{2n}-\tau_i)\backslash D_{2n-1}^i|, $$
and then  integrate out the imaginary time $\tau_{2n}$ The result is
\begin{eqnarray}
&& (F-F_0)_{\rm imp}
 =\frac{1}{\beta}\sum_{\omega_n}\int_0^U \tilde{\chi}_{\rm odd}(\omega_n)dU
\nonumber \\ &&
 =\frac{2}{\beta^3}\int_0^U dU \sum_{\omega_1,\omega_2,\omega_n}
  G_{\rm sc}(\omega_1)G_{\rm vec}(\omega_2)G_{\rm vec}(\omega_n-\omega_1)
       G_{\rm vec}(\omega_n+\omega_2)
 \nonumber \\ && \hspace{2cm}
 \Gamma_{0,1,2,3}(\omega_1,\omega_2;\omega_n-\omega_1,-\omega_n-\omega_2).
\end{eqnarray}

\subsection{Local conduction electron dynamical susceptibilities and singlet 
             superconducting pairing correlation function}

The conduction electron charge and spin density operators are defined in terms 
of the corresponding Majorana fermions,
\begin{eqnarray}
&& S_c^z(n)=\frac{1}{2}\left [C^{\dag}_{\uparrow}(n)C_{\uparrow}(n)
   -C^{\dag}_{\downarrow}(n)C_{\downarrow}(n)\right ]
  =-\frac{i}{2}\left[\Psi_1(n)\Psi_2(n)-\Psi_0(n)\Psi_3(n)\right];
\nonumber\\ &&
   n_c(n)=\frac{1}{2}\left [C^{\dag}_{\uparrow}(n)C_{\uparrow}(n)
   +C^{\dag}_{\downarrow}(n)C_{\downarrow}(n)-1 \right ]
  =-\frac{i}{2}\left[\Psi_1(n)\Psi_2(n)+\Psi_0(n)\Psi_3(n)\right],
\end{eqnarray}
and their density-density correlation functions at the impurity site are given
by
\begin{eqnarray}
&& \langle T_{\tau}S^z_c(0,\tau)S_c^z(0,\tau')\rangle
 =\langle T_{\tau}n_c(0,\tau)n_c(0,\tau')\rangle
\nonumber \\ && 
 =\frac{1}{4}\left \{ 
  \langle T_{\tau}\Psi_1(0,\tau)\Psi_1(0,\tau')
          \Psi_2(0,\tau)\Psi_2(0,\tau')\rangle
 +\langle T_{\tau}\Psi_0(0,\tau)\Psi_0(0,\tau')\rangle
         \langle T_{\tau}\Psi_3(0,\tau)\Psi_3(0,\tau')\rangle \right\},
\end{eqnarray}
where the conduction electron scalar operators decouple from the vector 
operators because they are free Majorana fermions in the  
model. Due to the hybridization 
between the impurity vector field and the conduction electron vector field,
the spectral function of these two density-density correlations can be derived 
from the perturbation expansion series of $U$ in the same way we did for the 
impurity density-density correlation functions. The final result is expressed as
\begin{eqnarray}\label{cchi}
 && \chi_{\rho,\sigma}^{\rm con}(\omega_n)
 =\frac{1}{4N^2}\sum_{k_1,k_2,k_3,k_4}
\left\{\frac{1}{\beta}\sum_{\omega_1} \left [G_{k_1,k_2}^{\alpha}(\omega_1)
  +G_{k_1,k_2}^0(\omega_1)\right]G_{k_3,k_4}^{\alpha}(\omega_n-\omega_1)
 \right.
 \nonumber \\ && \hspace{.5cm} 
  -\frac{1}{\beta^2}\sum_{\omega_1,\omega_2}
   G_{k_1,d}(\omega_1)G_{k_2,d}(\omega_2)G_{k_3,d}(\omega_n-\omega_1)
      G_{k_4,d}(\omega_n+\omega_2)
\Gamma'_{1,1,2,2}(\omega_1,\omega_2;
          \omega_n-\omega_1,-\omega_n-\omega_2)
\nonumber \\ && \hspace{.5cm}
-\frac{1}{\beta}\sum_{\omega_1}G_{k_1,d}(\omega_1)G_{k_2,d}(\omega_1)
      G_{k_3,k_4}^{\alpha}(\omega_n-\omega_1)
 \Sigma'_{\rm vec}(\omega_1)
\nonumber \\ && \hspace{.5cm}
  \left. -\frac{1}{\beta}\sum_{\omega_1}
  G_{k_3,d}(\omega_1)G_{k_4,d}(\omega_1)
       \left [G_{k_1,k_2}^0(\omega_n-\omega_1)
         +G_{k_1,k_2}^{\alpha}(\omega_n-\omega_1)\right]
    \Sigma'_{\rm vec}(\omega_1)\right\}. 
\end{eqnarray}

Meanwhile, we can consider the singlet superconducting pairing 
correlation function at the impurity site which is defined by  
\begin{equation}
 \chi_{\rm sup}^{\rm con}(\omega_n)=
 \frac{1}{\beta}\int_0^{\beta}d\tau\int_0^{\beta}d\tau'
  \langle T_{\tau}C_{\downarrow}(0,\tau)C_{\uparrow}(0,\tau)
         C^{\dag}_{\uparrow}(0,\tau')C^{\dag}_{\downarrow}(0,\tau')
   \rangle e^{i\omega_n(\tau-\tau')},
\end{equation}
where $\omega_n=2n\pi/\beta$. Using the Majorana fermion 
representation, we obtain the following expression 
\begin{eqnarray}
&& \langle T_{\tau}C_{\downarrow}(0,\tau)C_{\uparrow}(0,\tau)
    C^{\dag}_{\uparrow}(0,\tau')C^{\dag}_{\downarrow}(0,\tau')\rangle
\nonumber \\ &&
=\frac{1}{4}\left \{
  \langle T_{\tau}\Psi_1(0,\tau)\Psi_1(0,\tau')
          \Psi_3(0,\tau)\Psi_3(0,\tau')\rangle
   +\langle T_{\tau}\Psi_2(0,\tau)\Psi_2(0,\tau')
          \Psi_3(0,\tau)\Psi_3(0,\tau')\rangle \right.
 \nonumber \\ && 
 \left. +\langle T_{\tau}\Psi_0(0,\tau)\Psi_0(0,\tau')\rangle
         \langle T_{\tau}\Psi_1(0,\tau)\Psi_1(0,\tau')\rangle 
 +\langle T_{\tau}\Psi_0(0,\tau)\Psi_0(0,\tau')\rangle
         \langle T_{\tau}\Psi_2(0,\tau)\Psi_2(0,\tau')\rangle 
    \right\},
\end{eqnarray}
and, a relation with $\chi_{\rho,\sigma}^{\rm con}$ is found 
\begin{equation}\label{sup}
 \chi_{\rm sup}^{\rm con}(\omega_n)
 =2\chi_{\rho,\sigma}^{\rm con}(\omega_n).
\end{equation}

For the present Anderson-type model, we have several general relations 
between the 
conduction electron propagators and the impurity propagators. These relations 
for the Majorana Green functions are
\begin{eqnarray}
 && G_{k_1,k_2}^{\alpha}(\omega_1)
  =\frac{\delta_{k_2,k_1}}{i\omega_1-\epsilon_{k_1}}
   +\frac{V^2}{N}\frac{G_{\alpha}(\omega_1)}
      {(i\omega_1-\epsilon_{k_1})(i\omega_1-\epsilon_{k_2})}; 
  \nonumber \\ &&
    G_{k_1,k_2}^0(\omega_1)
    =\frac{\delta_{k_2,k_1}}{i\omega_1-\epsilon_{k_1}};
 \nonumber \\&& 
  G_{k_1,d}(\omega_1)
    =\frac{iV}{\sqrt{N}}
   \frac{G_{\alpha}(\omega_1)}{i\omega_1-\epsilon_{k_1}}.
\end{eqnarray}
Substituting these relations into the expression (\ref{cchi}), we finally get
the result
\begin{eqnarray}
 && \chi_{\rho,\sigma}^{\rm con}(\omega_n)
  -\chi_{\rho,\sigma}^{\rm con. (0)}(\omega_n)
\nonumber \\ &&
 =i\frac{3V^2}{4\beta}(\pi\rho)^3\sum_{\omega_1}G_{\alpha}^2(\omega_1)
      \Sigma'_{\rm vec}(\omega_1){\rm sgn}(\omega_n-\omega_1)
  +\frac{V^4}{2\beta}(\pi\rho)^4\sum_{\omega_1}G_{\alpha}^2(\omega_1)
     G_{\alpha}(\omega_n-\omega_1)\Sigma'_{\rm vec}(\omega_1) 
 \nonumber \\ && \hspace{.5cm}
 -\frac{V^4}{4\beta^2}(\pi\rho)^4\sum_{\omega_1,\omega_2}
    G_{\alpha}(\omega_1)G_{\alpha}(\omega_2)G_{\alpha}(\omega_n-\omega_1)
      G_{\alpha}(\omega_n+\omega_2)
  \nonumber \\ && \hspace{1.5cm}
  \Gamma'_{1,1,2,2}(\omega_1'\omega_2;\omega_n-\omega_1,-\omega_n-\omega_2) 
 {\rm sgn \omega_1}{\rm sgn \omega_2}{\rm sgn (\omega_n-\omega_1)}
     {\rm sgn (\omega_n+\omega_2)},
\end{eqnarray}
where the density-density spectra in the case of $U=0$ are 
\begin{eqnarray}
&& \chi_{\rho,\sigma}^{\rm con. (0)}(\omega_n)
  =\frac{1}{2\beta}(\pi\rho)^2\sum_{\omega_1}{\rm sgn}\omega_1
    {\rm sgn}(\omega_n-\omega_1) 
  +i\frac{3V^2}{4\beta}(\pi\rho)^3\sum_{\omega_1}G_{\alpha}(\omega_1)
     {\rm sgn}(\omega_n-\omega_1)
\nonumber \\ && \hspace{3cm}
  + \frac{V^4}{4\beta}(\pi\rho)^4\sum_{\omega_1}G_{\alpha}(\omega_1)
      G_{\alpha}(\omega_n-\omega_1),
\end{eqnarray}
which is a regular contribution. When the Matsubara frequency is taken to zero,
 the corresponding static susceptibilities are also obtained.  
With the help of the relation (\ref{sup}), the conduction electron singlet 
superconducting pairing correlation function at the impurity site can also be 
obtained.

\section{Main perturbational results of the model}

\subsection{Leading specific heat correction}
  
In the  section IV, an expression for
the impurity free energy was derived in a power series in 
$U$. Since $F^{(0)}_{\rm imp}$ is the corresponding free energy in the case of 
$U=0$ and is just a regular contribution, the first singular contribution 
to the free energy comes in second order in $U$, and is
\begin{eqnarray}
&& F^{(2)}_{\rm imp}=-\frac{U^2}{4\beta}\int_0^{\beta}d\tau_2
   \int_0^{\beta}d\tau_1 G_{\alpha}^3(\tau_1-\tau_2){\rm sgn}(\tau_2-\tau_1)
 \nonumber \\ && \hspace{1cm}
 =\frac{U^2}{2\beta^3}\sum_{\omega_1,\omega_2,\omega_3}
  G_{\alpha}(\omega_1)G_{\alpha}(\omega_2)G_0(\omega_3)
    G_{\alpha}(\omega_1+\omega_2+\omega_3).
\end{eqnarray}
The singular contribution from this term is
$$ F^{(2)}_{\rm imp}\approx\frac{1}{4}\left (\frac{U}{\pi\Delta}\right )^2
  \frac{(\pi T)^2}{\pi\Delta}\ln\left(\frac{\Delta}{T}\right),$$  
which gives a singular correction to the impurity specific heat,   
$$ C^{(2)}_{\rm imp}\approx\frac{\pi^2}{2}\left(\frac{U}{\pi\Delta}\right)^2
\frac{T}{\pi\Delta}\ln\left(\frac{\Delta}{T}\right).$$ 

\subsection{Self energies and electrical resistivity}

From the general expressions for the self-energies, it can be seen that
the lowest order self
energy corrections are of  second order. For the impurity scalar field, the 
perturbed self energy is
\begin{eqnarray}
&& \Sigma^{(2)}_{\rm sc}(\omega_n)
   = - 2i\frac{U^2}{\beta}\int_0^{\beta}d\tau_2\int_0^{\tau_2}d\tau_1
      G_{\alpha}^3(\tau_1-\tau_2)\sin[\omega_n(\tau_1-\tau_2)]
\nonumber \\ && \hspace{2cm}
  = -\frac{U^2}{2}\int_{-\beta}^{\beta}G_{\alpha}^3(\tau)
                 e^{i\omega_n\tau}d\tau.
\end{eqnarray}
In the imaginary time space, this self energy is 
\begin{equation}
 \Sigma^{(2)}_{\rm sc}(\tau)=-U^2G_{\alpha}^3(\tau),
\end{equation}
which can be represented by a Feynman diagram shown in Fig.3a, and its 
Fourier transform is given by
\begin{equation}
 \Sigma^{(2)}_{\rm sc}(\omega_n)
 =-\frac{U^2}{\beta^2}\sum_{\omega_1,\omega_2}G_{\alpha}(\omega_1)
            G_{\alpha}(\omega_2)G_{\alpha}(\omega_n-\omega_1-\omega_2).
\end{equation}  
The imaginary part of its retarded Fourier transform after integration is
given by
\begin{equation}
 {\rm Im}\Sigma^{(2)}_{\rm sc}(\omega,T)
  \approx -\frac{\Delta}{2}\left(\frac{U}{\pi\Delta}\right)^2 
                \left[\left(\frac{\omega}{\Delta}\right)^2
                       +\left(\frac{\pi T}{\Delta}\right)^2\right].
\end{equation}
Here the analytical continuation has been performed, and according to the 
Kramers-Kronig relation the real part should be
\begin{equation}
 {\rm Re}\Sigma^{(2)}_{\rm sc}(\omega,T)
   \approx -\left(\frac{U}{\pi\Delta}\right)^2\omega.
\end{equation}
Moreover, the renormalization factor of the wave function or quasiparticle 
weight ($z_{\rm sc}(\omega_n)$ taken at $\omega_n=0$) to second order is given 
by 
\begin{equation}
  z_{\rm sc}^{(2)}=\left[1-\frac{\partial{\rm Re}\Sigma_{\rm sc}^{(2)}}
                                {\partial\omega}\right]^{-1}_{\omega=0}
                  =\left [1+\left(\frac{U}{\pi\Delta}\right)^2 \right]^{-1}.
\end{equation}
The scalar self energy exhibits the normal local FL behavior, and such a 
behavior ensures the fermionic zero mode in the unperturbed Hamiltonian is 
preserved in the presence of perturbations.

On the other hand, the vector perturbed self energy can be written down
from the general expression,
\begin{eqnarray}
&& \Sigma^{(2)}_{\rm vec}(\omega_n)
 = -i \frac{U^2}{\beta}\int_0^{\beta}d\tau_2\int_0^{\tau_2}d\tau_1
 G_{\alpha}^2(\tau_1-\tau_2)\sin[\omega_n(\tau_1-\tau_2)]
\nonumber \\ && \hspace{2cm}
 = \frac{U^2}{4}\int_{-\beta}^{\beta}d\tau G_{\alpha}^2(\tau){\rm sgn}(\tau)
  e^{i\omega_n\tau}.
\end{eqnarray}
The sign function appears when the double integrals are
reduced to a single integral. In the same way, the self energy in imaginary 
time space becomes
\begin{equation}
 \Sigma^{(2)}_{\rm vec}(\tau)=-U^2G_{\alpha}^2(\tau)G_0(\tau),
\end{equation}
which corresponds to the Feynman diagram in Fig.3b, and the Fourier
transform is given by
\begin{equation}
 \Sigma^{(2)}_{\rm vec}(\omega_n)
 =-\frac{U^2}{\beta^2}\sum_{\omega_1,\omega_2}G_{\alpha}(\omega_1)
            G_{\alpha}(\omega_2)G_0(\omega_n-\omega_1-\omega_2).
\end{equation}
Since this self energy involves the fermionic zero mode, the calculations of 
its spectral function becomes subtle. The retarded imaginary part is given by
\begin{equation}\label{imvec}
 {\rm Im}\Sigma^{(2)}_{\rm vec}(\omega,T)
 =-\frac{\pi}{2}\left(\frac{U}{\pi\Delta}\right)^2|\omega|
            {\rm coth}\left(\frac{|\omega|}{2T}\right)
\sim\left\{
\begin{array}{ll}
 -(\frac{U}{\pi\Delta})^2(\pi T), & |\omega|\ll T\\
 -\frac{\pi}{2}(\frac{U}{\pi\Delta})^2|\omega|, & 
\Delta\gg |\omega|\gg T. 
\end{array}
\right.
\end{equation}
The corresponding real part is also obtained as
\begin{equation}
 {\rm Re}\Sigma^{(2)}_{\rm vec}(\omega,T)
   \approx \left(\frac{U}{\pi\Delta}\right)^2
           \omega\ln\left(\frac{x}{\Delta}\right),  
\end{equation}
where $x={\rm max}(|\omega|,T)$.
Such a self energy is greatly different from the form given by 
the ordinary FL theory, in particular, the renormalization factor of the wave 
function or the quasiparticle weight logarithmically approaches to zero as
$T\rightarrow 0$,   
\begin{equation}
 z_{\rm vec}^{(2)}
 =\left[1-\frac{\partial{\rm Re}\Sigma_{\rm vec}^{(2)}}
               {\partial\omega}\right]^{-1}_{\omega=0}
 =\left[1+\left(\frac{U}{\pi\Delta}\right)^2
       \ln\left(\frac{\Delta}{T}\right)\right]^{-1},
\end{equation}
which implies that the impurity vector self-energy displays the marginal 
FL behavior. 

There is a singular temperature-dependent contribution to
${\rm Im}\Sigma^{(2)}_{\rm vec}(0,T)$,
and the conduction electron t-matrix determined by the Green function of
the impurity vector propagator has been derived as follows  
\begin{equation}
 t(\omega,T)=\frac{V^2}{N}G_{\rm vec}(\omega,T).
\end{equation}
 Thus, the imaginary part of this t-matrix can be expressed in terms of the
retarded self energy for the impurity vector field:
\begin{eqnarray}
 && {\rm Im}t(\omega, T)
\nonumber \\ &&
= -\frac{1}{\pi\rho N}\frac{\Delta^2}
      {(\omega-{\rm Re}\Sigma_{\rm vec})^2+(\Delta-{\rm Im}\Sigma_{\rm vec})^2}
      + \frac{\Delta}{\pi\rho N}\frac{{\rm Im}\Sigma_{\rm vec}}
 {(\omega-{\rm Re}\Sigma_{\rm vec})^2+(\Delta-{\rm Im}\Sigma_{\rm vec})^2},
\end{eqnarray}
where the first term describes the elastic scatterings, while the
second term describes the inelastic scatterings.

Assuming the conduction electrons incoherently scatter from the dilute
magnetic impurities, the linear response theory allows the
electrical conductivity to be expressed in terms of the following Kubo formula
\begin{equation}
 \sigma(T)=-\frac{2}{3}e^2v_f^2\rho\int_{-\infty}^{\infty}
             \tau(\omega,T)\frac{\partial f}{\partial \omega}d\omega,
\end{equation}
here $f(x)$ is the Fermi distribution function, $v_f$ is the Fermi velocity of
the conduction electrons with charge $e$ and density of states $\rho$, and
$\tau(\omega, T)$ is the electron relaxation time, which is related to the
t-matrix
$$ \tau^{-1}=-2N_{\rm imp}{\rm Im}t(\omega,T) $$
where $N_{\rm imp}$ is the total number of the impurities, and
$n_{\rm imp}=N_{\rm imp}/N$ is the impurity concentration and is supposed to
be much less than unity. On substituting the second order self energy for the
impurity vector field into equations (82) and (83), we  obtain the electrical
 resistivity in the low temperature regime,
\begin{equation}
  \rho(T)\approx\frac{3\pi n_{\rm imp}}{e^2}
     \left[ 1 + \left(\frac{U}{\pi\Delta}\right)^2
     \left(\frac{\pi T}{\Delta}\right)\right ].
\end{equation}
Here we have just taken into account the contributions of
 order  $(\frac{U}{\pi\Delta})^2$.
Therefore, up to the second order of perturbation series, the electrical 
resistivity has a linear temperature dependence, which is a direct consequence 
of 
the anomalous behavior of the imaginary part of the impurity vector fermion 
self energy. {\it Such a resistivity makes the weak coupling fixed point of 
the present model differ from the strong coupling fixed point of the 
two-channel Kondo model}.

\subsection{Vertex functions in the zero-frequency limit}

As we said earlier, in the zero-frequency limit there is only one non-zero 
two-particle vertex function $\Gamma'_{0,1,2,3}$, the other two vertex 
functions vanish. Its general expression in terms of $U$ has been
given in section IV. In the first order perturbation, it is trivial to
obtain as
\begin{equation}
  \Gamma'^{(1)}_{0,1,2,3}(\omega,\omega',\omega'',\omega''')
  =-U\delta_{\omega+\omega'+\omega''+\omega''',0},
\end{equation}
the corresponding diagram is described in Fig.4a.  
To order $U^3$, it is given by
\begin{eqnarray}
&& \Gamma'^{(3)}_{0,1,2,3}(0,0,0,0)=-\frac{U^3}{12\beta}
\int_0^{\beta}d\tau_3\int_0^{\beta}d\tau_2\int_0^{\beta}d\tau_1
\left[G_{\alpha}(\tau_1-\tau_2)-G_{\alpha}(\tau_1-\tau_3)+
   G_{\alpha}(\tau_2-\tau_3)\right]^3_l 
\nonumber \\ && \hspace{6cm}
  {\rm sgn}(\tau_2-\tau_1){\rm sgn}(\tau_3-\tau_2){\rm sgn}(\tau_3-\tau_1).
\end{eqnarray}
Using the diagrammatic method, we can clarify that four basic Feynman diagrams 
have contributions to this zero-frequency vertex function, and then re-group
these terms to get the following expression,   
\begin{eqnarray}
&& \Gamma'^{(3)}_{0,1,2,3}(0,0,0,0)=\frac{U^3}{2\beta}
 \int_0^{\beta}d\tau_3\int_0^{\beta}d\tau_2\int_0^{\beta}d\tau_1
 \{ 3 G_{\alpha}^2(\tau_1-\tau_2)G_{\alpha}(\tau_1-\tau_3)
   {\rm sgn}(\tau_1-\tau_3)
 \nonumber \\ && \hspace{1cm}
 -G_{\alpha}^2(\tau_1-\tau_2)G_{\alpha}(\tau_1-\tau_3){\rm sgn}(\tau_1-\tau_2)
 -G_{\alpha}^2(\tau_1-\tau_2)G_{\alpha}(\tau_1-\tau_3){\rm sgn}(\tau_2-\tau_3)
\nonumber \\ && \hspace{1cm}
 -G_{\alpha}^2(\tau_1-\tau_2)G_{\alpha}(\tau_1-\tau_3)G_{\alpha}(\tau_2-\tau_3)
    {\rm sgn}(\tau_1-\tau_2) \},
\end{eqnarray}
where each term is described by the respective Feynman diagram (b), (c), (d), 
(e)
in Fig.4, and it is easily found that only the first term (Fig.4b) gives the 
leading singular contributions,  while the remaining three terms can be 
neglected in the limit $T\rightarrow 0$. Thus, the leading third order 
perturbation correction to the vertex function is   
\begin{eqnarray}
&&  \Gamma'^{(3)}_{0,1,2,3}(0,0,0,0)
 =-3 U^3\left [-\frac{1}{\beta}\sum_{\omega_n}G^2(\omega_n) \right ]
  \left [-\frac{1}{\beta}\sum_{\omega_n}G(\omega_n)G_0(\omega_n) \right]
 \nonumber \\ && \hspace{3cm}
 \approx -3U\left(\frac{U}{\pi\Delta}\right)^2\ln\left(\frac{\Delta}{T}\right),
    \hspace{3cm} {\rm for} \hspace{.5cm} T\ll\Delta, 
\end{eqnarray}
and a logarithmic singularity appears. Up to  third order in $U$, the 
proper vertex function is calculated and exhibits the same singularity as well, 
\begin{equation}
 \Gamma_{0,1,2,3}(0,0,0,0)
 =z_{\rm sc}^{-1}z_{\rm vec}^{-3}\Gamma'_{0,1,2,3}(0,0,0,0)
 \approx -U\left\{1+\left(\frac{U}{\pi\Delta}\right)^2
           +6\left(\frac{U}{\pi\Delta}\right)^2
             \ln\left(\frac{\Delta}{T}\right)\right\},
\end{equation}
which implies that the higher order perturbative expansion terms will become 
as important as low-order contributions in determining the low-temperature 
behavior of the model. At least the leading order logarithmic terms have to be 
summed to obtain the correct low-temperature behavior. The summation of 
all the leading logarithmic contributions for the vertex function is the 
so-called parquet approximation. In section VII, we will use another equivalent 
approach: the multiplicative renormalization-group method, to find the final 
results. 

\subsection{Impurity static susceptibilities}

We have already obtained  general expressions for the impurity dynamical 
susceptibilities in terms of the impurity scalar and vector self energies and 
their two-particle vertex functions. Since these associated self energies and 
vertex functions have been  obtained in the lowest order perturbation theory, 
we can use them to derive the even and odd static susceptibilities. 
In the zeroth order, there are two contributions, 
\begin{equation}
  \tilde{\chi}^{(0)}_{\rm even}=
 \left [-\frac{1}{\beta}\sum_{\omega_n}G^2_{\alpha}(\omega_n)\right ]+
\left [-\frac{1}{\beta}\sum_{\omega_n}G_{\alpha}(\omega_n)G_0(\omega_n)\right],
\end{equation}
where the first term is a regular FL like contribution, while the second term 
shown in Fig.5a is singular, and can be expressed as
\begin{equation}
 \left [ -\frac{1}{\beta}\sum_{\omega_n}G_{\alpha}(\omega_n)
           G_0(\omega_n) \right ]
 = \left ( \frac{1}{\pi\Delta}\right )
\left [\psi\left(\frac{1}{2}+\frac{\Delta}{2\pi T}\right) -
  \psi\left(\frac{1}{2}\right)\right],
\end{equation}
where $\psi(x)$ is the digamma function, and when $T\ll\Delta$, 
 $ \left [\psi\left(\frac{1}{2}+\frac{\Delta}{2\pi T}\right) -
  \psi\left(\frac{1}{2}\right)\right]\approx\ln\left(\frac{\Delta}{T}\right)$,
and $\left [-\frac{1}{\beta}\sum_{\omega_n}G^2_{\alpha}(\omega_n)\right] 
     \approx  \left ( \frac{1}{\pi\Delta}\right )$.
Thus, the singular contribution gives rise to a logarithmic divergence in 
the impurity static susceptibilities. In the first order in $U$, there is also 
one singular term 
corresponding to the diagram in Fig.5b. From the first-order vertex correction 
for $\Gamma'_{0,1,2,3}(\omega_1,\omega_2;-\omega_1,-\omega_2)$, we get  
\begin{equation}
   \tilde{\chi}^{(1)}_{\rm odd}
 = 2U \left [-\frac{1}{\beta}\sum_{\omega_n}G^2(\omega_n)\right ]
     \left [-\frac{1}{\beta}\sum_{\omega_n}G(\omega_n)G_0(\omega_n) \right].
\end{equation}
While to second order in $U$, the perturbative corrections to the impurity
static susceptibilities become complicated, 
\begin{eqnarray}
 && \tilde{\chi}^{(2)}_{\rm even}
 = U^2 \left [-\frac{1}{\beta}\sum_{\omega_n}G^2(\omega_n)\right ]^2
     \left [-\frac{1}{\beta}\sum_{\omega_n}G(\omega_n)G_0(\omega_n)\right]
\nonumber \\ && \hspace{1.5cm}
 +U^2 \left [-\frac{1}{\beta}\sum_{\omega_n}G^2(\omega_n)\right ]
    \left [-\frac{1}{\beta}\sum_{\omega_n}G(\omega_n)G_0(\omega_n)\right]^2
  +{\rm less \hspace{.1cm} singular \hspace{.1cm} terms},
\end{eqnarray} 
where we have used the second-order results for the impurity self-energies 
$\Sigma'^{(2)}_{\rm sc}(\omega_n)$ and $\Sigma'^{(2)}_{\rm vec}(\omega_n)$ and
vertex functions 
$\Gamma'^{(2)}_{1,1,2,2}(\omega_1,\omega_2;-\omega_1,-\omega_2)$ and 
$\Gamma'^{(2)}_{0,0,3,3}(\omega_1,\omega_2;-\omega_1,-\omega_2)$.
Note that apart from the logarithmic contributions represented by the diagram 
in Fig.5c, there appears another singular term: a squared logarithmic term 
arising from  diagram (d) in Fig.5.
Therefore, up to the second order in the perturbation series, the singular 
parts of the impurity spin and charge static susceptibilities are given by
\begin{eqnarray}
&& \chi^{\rm imp}_{\sigma}
 \approx \left(\frac{g\mu_B}{2}\right)^2 \left\{ 
       \left[1+ 2\left(\frac{U}{\pi\Delta}\right)
             +\left(\frac{U}{\pi\Delta}\right)^2 \right]
       \left(\frac{1}{\pi\Delta}\right)\ln\left(\frac{\Delta}{T}\right)
     +\left(\frac{U}{\pi\Delta}\right)^2\left(\frac{1}{\pi\Delta}\right)
       \ln^2\left(\frac{\Delta}{T}\right) \right\}, 
\nonumber \\ && 
\chi^{\rm imp}_c
\approx \frac{1}{4} \left \{
    \left[1- 2\left(\frac{U}{\pi\Delta}\right)
          + \left(\frac{U}{\pi\Delta}\right)^2 \right]
    \left(\frac{1}{\pi\Delta}\right)\ln\left(\frac{\Delta}{T}\right)
  +\left(\frac{U}{\pi\Delta}\right)^2\left(\frac{1}{\pi\Delta}\right)
       \ln^2\left(\frac{\Delta}{T}\right) \right\}.
\end{eqnarray}

\subsection{Conduction electron local static susceptibilities and singlet 
            superconducting pairing susceptibility}
 
For the local static susceptibilities of the conduction electrons, it is 
straight forward to obtain  results for the local static susceptibilities 
of the conduction electrons by taking the zero-frequency limit
for the dynamical susceptibilities. As we have stated before, 
the unperturbed charge and spin static susceptibilities of the conduction
electrons at the impurity site are regular. Then, after subtracting the 
contribution of the free conduction electrons ($V=0$), it is expressed as 
\begin{equation}
  \delta\chi_{\rho,\sigma}^{{\rm con}. (0)}
 \approx \frac{1}{4}(\pi\rho V)^4
    \left[-\frac{1}{\beta}\sum_{\omega_n}G_{\alpha}^2(\omega_n)\right]
 \approx \frac{1}{4}(\pi\rho V)^4\left(\frac{1}{\pi\Delta}\right).
\end{equation}
Unlike the impurity static susceptibilities, the first perturbative 
correction comes from second order in $U$.  Since the vertex correction 
$\Gamma'^{(2)}_{1,1,2,2}(\omega_1,\omega_2;
  \omega_n-\omega_1,-\omega_n-\omega_2)$ involves a singular 
contribution analogous to the diagram Fig.5c, the singular part of the 
second-order correction is 
\begin{equation}
  \delta\chi_{\rho,\sigma}^{{\rm con}. (2)}
   =-\frac{U^2}{4}(\pi\rho V)^4 
    \left[-\frac{1}{\beta}\sum_{\omega_n}G_{\alpha}^2(\omega_n)\right]^2
 \left[-\frac{1}{\beta}\sum_{\omega_n}G_{\alpha}(\omega_n)G_0(\omega_n)\right].
\end{equation}
When $T\ll \Delta$, up to the second order in $U$, the static susceptibilities 
are given by
\begin{equation}
 \delta\chi_{\rho,\sigma}^{\rm con} 
 \approx \frac{1}{4}(\pi\rho V)^4 \left(\frac{1}{\pi\Delta} \right )
        -\frac{1}{4}(\pi\rho V)^4\left(\frac{U}{\pi\Delta}\right)^2 
     \left (\frac{1}{\pi\Delta}\right)\ln \left (\frac{\Delta}{T} \right), 
\end{equation}
where there appears a logarithmically temperature dependent contribution. 
The possible reason is that the impurity vector field transfers its 
low-energy singularity into the local conduction electrons through the 
hybridization, so there is also a logarithmic temperature dependence of the 
impurity static susceptibilities to second order in $U$. 

Another important result is the second order correction to the singlet 
superconducting pairing correlation function at the impurity site, its static 
susceptibility can also  be  obtained through the relation 
$\chi_{\rm sup}^{\rm con}=2\chi^{\rm imp}_{\rho,\sigma}$,
\begin{equation}
 \delta\chi_{\rm sup}^{\rm con}
 \approx  \frac{1}{2}(\pi\rho V)^4\left(\frac{1}{\pi\Delta}\right)
       -\frac{1}{2}(\pi\rho V)^4\left(\frac{U}{\pi\Delta}\right)^2
      \left (\frac{1}{\pi\Delta}\right)\ln\left (\frac{\Delta}{T} \right),
\end{equation}
which indicates that the conduction electrons form a singlet pairing resonant 
state around the local impurity site, different 
from a single-particle resonant state in the single-channel Kondo problem. 
It should be pointed out that this pairing resonance is pinned at the Fermi 
level (chemical potential).
 
\section{Multiplicative renormalization-group analysis}

Since the perturbation theory on the present single-impurity model gives
logarithmic contributions to the impurity vertex function and vector field
propagator, the summation of the leading order logarithmic terms
should lead to some form of scaling behavior. One way to sum these terms is by 
using the usual parquet approximation. Here we use another equivalent method 
to extend our second order perturbational results for the impurity self 
energies and vertex function: the multiplicative renormalization-group (RG) 
\cite{bs}. 

The usual multiplicative RG is a simple transformation procedure in 
which the Green functions, vertices and coupling constants are multiplied by 
real, frequency independent factors. The requirement that the two-particle 
interaction form defined in the model Hamiltonian be satisfied by both the 
original and the transformed quantities  gives a relation between them. The 
classical formulation was used to obtain scaling laws for the X-ray edge 
problem and the single-channel Kondo problem \cite{amfz,solyom}. For the 
logarithmic problem, generally, 
it is important to give a proper definition of the invariant coupling,
the temperature dependence of which characterizes the behavior of the system. 
On the physical ground, we take the hybridization width $\Delta$ as a 
natural scaling parameter. The perturbed impurity Green functions and the 
important two-particle vertex function are written in the form,
\begin{eqnarray}
  && G_{\rm sc}(\omega)=z_{\rm sc}G_0(\omega),
     \hspace{4cm}
     G_{\rm vec}(\omega)=z_{\rm vec}G_{\alpha}(\omega);
\nonumber \\ &&
      \bar{U}\tilde{\Gamma}_{0,1,2,3}
      \equiv -\left(\frac{1}{\pi\Delta}\right)\Gamma_{0,1,2,3}(0,0,0,0), 
   \hspace{1cm}
      \bar{U}\equiv\left(\frac{U}{\pi\Delta}\right),
\end{eqnarray}
where $G_0$ and $G_{\alpha}$ are the unperturbed Green functions and 
$\bar{U}$ is the dimensionless bare vertex. For simplicity, in the functions 
of $z_{\rm sc}$, $z_{\rm vec}$, and $\tilde{\Gamma}_{0,1,2,3}$, the frequency 
variables are fixed at the Fermi level and only the temperature variables are 
retained. If the interaction is cut off at the energy $\Delta$, the Green 
functions and vertex depend, as a rule, on the relative energies $\Delta/T$.

Multiplicative RG is usually formulated as the transformation induced by a 
change of 
the cut-off parameter from $\Delta$ to $\Delta'$, $i.e.$.
\begin{eqnarray}
  && z_{\rm sc}\left(\frac{\Delta'}{T},\bar{U'}\right)
      = z_1 z_{\rm sc}\left(\frac{\Delta}{T},\bar{U}\right), 
 \nonumber \\ 
   && z_{\rm vec}\left(\frac{\Delta'}{T},\bar{U'}\right)
       =z_2 z_{\rm vec}\left(\frac{\Delta}{T},\bar{U}\right), 
 \nonumber \\
   && \tilde{\Gamma}_{0,1,2,3}\left(\frac{\Delta'}{T},\bar{U'}\right)
       =z_3^{-1} \tilde{\Gamma}_{0,1,2,3}
           \left(\frac{\Delta}{T},\bar{U}\right),
 \nonumber \\
   && \bar{U'}\left(\frac{\Delta'}{\Delta}\right) 
       =z_1^{-1/2}z_2^{-3/2}z_3\bar{U},
\end{eqnarray}
where $z_1$, $z_2$, and $z_3$ are independent of the temperature 
variables. 
Such a transformation can keep the four-legged vertex unchanged, which 
corresponds to the two-particle interaction of the model Hamiltonian. Since 
$ z_{\rm sc}^{(2)}=[1+\bar{U}^2]^{-1}$ does not depends on the cut-off factor
at all, it can not induce any essential renormalization. For simplicity, we 
choose $z_1=1$. Therefore, when the above relations are obeyed, the cut-off 
dependent $\bar{U'}$, a self-consistent solution to these equations is
given by
\begin{equation}
 \bar{U'}\left(\frac{\Delta'}{\Delta}\right)
  \tilde{\Gamma}_{0,1,2,3}\left(\frac{\Delta'}{T},\bar{U'}\right)
  \left[z_{\rm vec}\left(\frac{\Delta'}{T},\bar{U'}\right)\right]^{3/2}
 =\bar{U}\tilde{\Gamma}_{0,1,2,3}\left(\frac{\Delta}{T},\bar{U}\right)
   \left[z_{\rm vec}\left(\frac{\Delta}{T},\bar{U}\right)\right]^{3/2},
\end{equation}
or 
\begin{equation}
    \bar{U'}\left(\frac{\Delta'}{\Delta}\right)
         =\bar{U}\frac{\tilde{\Gamma}_{0,1,2,3}(\frac{\Delta}{T},\bar{U})}
                {\tilde{\Gamma}_{0,1,2,3}(\frac{\Delta'}{T},\bar{U'})}
           \frac{\left[z_{\rm vec}(\frac{\Delta}{T},\bar{U})\right]^{3/2}}
             {\left[z_{\rm vec}(\frac{\Delta'}{T},\bar{U'})\right]^{3/2}},
\end{equation}
which is the invariant coupling constant of this model. When the denominator is 
normalized to unity at $T=\Delta'$, a simple definition of the invariant 
coupling constant is 
\begin{equation}
  \bar{U}_{\rm inv}
  =\bar{U}\tilde{\Gamma}_{0,1,2,3}
        \left(\frac{\Delta}{\Delta'},\bar{U}\right)
  \left[z_{\rm vec}\left(\frac{\Delta}{\Delta'},\bar{U}\right)\right]^{3/2}.
\end{equation}
   
Up to the second order in $\bar{U}$, the impurity renormalization factors of 
the wave functions and the vertex function have already been obtained as 
\begin{eqnarray}
  && z_{\rm vec}^{(2)}\left(\frac{\Delta}{T},\bar{U}\right)
     =\frac{1} {1+ \bar{U}^2\ln \left (\frac{\Delta}{T}\right)},
 \nonumber \\ 
  && \tilde{\Gamma}_{0,1,2,3}\left(\frac{\Delta}{T},\bar{U}\right)
     =1+\bar{U}^2+6\bar{U}^2\ln\left(\frac{\Delta}{T}\right).
\end{eqnarray}
Actually, the non-logarithmic parts are not essential in the present RG 
approach. From the above second-order results, the multiplicative factors $z_2$
and $z_3$ can be derived 
\begin{equation}
 z_2=\frac{1}
          { 1 + \bar{U}^2 \ln \left ( \frac{\Delta'}{\Delta} \right) },
 \hspace{.5cm}
 z_3=\frac{1}{1+6\bar{U}^2\ln\left(\frac{\Delta'}{\Delta}\right)},
\end{equation}  
and then the dimensionless invariant coupling is given by 
\begin{equation}
   \bar{U}'=\bar{U}\left [1-\frac{9}{2}\bar{U}^2
                        \ln\left(\frac{\Delta'}{\Delta}\right)+.....\right],
\end{equation}
from which the basic RG scaling equation can be cast into a differential form,
\begin{equation}\label{RG1}
  \frac{d\ln\bar{U}}{d\ln\Delta} =-\frac{9}{2}\bar{U}^2.
\end{equation}
The characteristic feature of this scaling equation is that the dimensionless 
coupling constant $\bar{U}=U/(\pi\Delta)$ always increases as the 
high-energy scale $\Delta$ is reduced. Integrating this differential equation 
gives the relation,
\begin{equation}
  \bar{U'}^2 =\frac{\bar{U}^2}
      {1+9\bar{U}^2 \ln\left( \frac{\Delta'}{\Delta} \right)}.
\end{equation}
The scaling relation can be used to extend the second-order perturbational 
results, and effectively sum the leading order logarithmic terms. The scaling
trajectory can be traversed from the initial parameters $\Delta$ and $U$, to
an effective $\Delta'$ of order $T$, and an effective coupling constant 
$\bar{U}'$, which becomes temperature dependent, 
$$ \bar{U}^2(T)=\frac{\bar{U}^2}
                    {1+9\bar{U}^2\ln\left(\frac{T}{\Delta}\right)}.$$
When we substitute this new temperature dependent coupling constant into the
second-order perturbational results, we will get the results which are 
equivalent to a summation of the leading
order logarithmic terms. For example, the electrical resistivity in the
second order perturbation theory was calculated to be  
\begin{equation}
 \rho^{(2)}(T)=\frac{3\pi n_{\rm imp}}{e^2}
     \left[1+\bar{U}^2\left(\frac{\pi T}{\Delta}\right)\right ].
\end{equation}
After substituting the new coupling parameter, it becomes
\begin{equation}
  \rho(T)=\frac{3\pi n_{\rm imp}}{e^2}\left[
      1+\frac{\bar{U}^2}{1+9\bar{U}^2\ln\left(\frac{T}{\Delta}\right)}
                    \left(\frac{\pi T}{\Delta}\right)\right ].
\end{equation}
The $\ln T$ term is in the denominator due to the summation of the leading
order logarithmic series. As $T\rightarrow 0$, the difficulties arising from 
the $\ln T$ term become more severe as there is a divergence at finite 
temperature,
\begin{equation}
T_c=\Delta{\rm exp}\left[-\frac{1}{9}\left(\frac{\pi\Delta}{U}\right)^2 \right],
\end{equation}
which is a new weak-coupling low-temperature energy scale.
When $T>T_c$, the perturbational scaling can be extended down to an effective
hybridization width $\Delta'$ and effective coupling constant $\bar{U}'$, 
while $T<T_c$, the electrical resistivity derived from perturbation scaling 
diverges.  

In the conventional RG treatments for the single-impurity Kondo problems 
\cite{anderson}, the conduction electron band-width $D$ is usually 
chosen as the high energy cut-off factor, and the coupling parameters are 
renormalized as the bandwidth $D$ is decreased. However, for the Anderson-type 
impurity model, the perturbation expansions in the interaction parameter $U$ by
Yamada and Yosida \cite{yy,yamada} and the numerical RG calculations 
\cite{wilson} 
have shown that the model provides its own {\it intrinsic} high-energy cut-off 
of the order of the hybridization width $\Delta$ ($\Delta > U$), inconsistent 
with using the conduction electron bandwidth as the effective cut-off factor 
\cite{haldane}. 
Similarly, the present perturbation results of the compactified Anderson 
impurity model have also proved that the parameter $\Delta$ 
plays the role of the high-energy cut-off factor. In fact, the behavior of the 
present model is independent of the conduction electron bandwidth, which can be
taken to the infinite bandwidth limit without lost of generality.  
It is noteworthy to point out that the scaling equation (\ref{RG1}) 
is a for the dimensionless coupling parameter $\bar{U}$. Due to 
$\bar{U}=U/(\pi\Delta)$, there is another form of the RG scaling in terms of 
the dimensional coupling parameter $U$, 
$ \frac{dU}{d\ln\Delta}
  =U\left[ 1-\frac{9}{2}\left(\frac{U}{\pi\Delta}\right)^2\right]$.
However, this form of the scaling equation is not useful for discussing the 
scaling behavior of the interaction because the parameter $U$ is not the 
RG invariant coupling parameter. 
The RG scaling of the model has explicitly displayed in Eq.(\ref{RG1}). 

What have we learned from these scaling arguments? First of all, the 
hybridization width $\Delta$ can be reduced dramatically for the calculation
of thermodynamic behavior from its initial value down to the thermal energy 
scale and still be described by a model of the same form but with a renormalized
coupling constant. Second, there is a subtle point in the perturbation 
treatments. Basically the model has been described by two independent coupling 
parameters $\Delta$ and $U$, and the perturbation expansion is based on the
following two-particle vertex function,
\begin{equation}
 \Gamma_{0,1,2,3}(0,0,0,0)=-U\tilde{\Gamma}_{0,1,2,3}(\bar{U}),
\end{equation} 
where $\tilde{\Gamma}_{0,1,2,3}$ is the corresponding dimensionless vertex 
function and only depends on the dimensionless coupling constant 
$\bar{U}=\frac{U}{\pi\Delta}$. The important point is that 
dimensionless coupling parameter $|\bar{U}|$ increases as the scaling 
parameter is reduced. $\bar{U}$ becomes infinite when the scaling parameter 
approaches the characteristic low-energy scale $T_c$, and the dimensionless 
vertex function diverges as well. Thus, the perturbation theory
begins to break down, and there is one weak-coupling low-energy scale $T_c$ 
depending on the dimensionless coupling parameter $\bar{U}$ only. 

\section{Discussions and conclusions}

From the previous perturbational calculations and the multiplicative RG 
analysis, we have found a non-FL behavior in the weak-coupling regime 
($\bar{U}\ll 1$), and that the invariant coupling parameter $\bar{U}$ 
increases as
the high-energy scale $\Delta$ is reduced. So our perturbation results are not
valid when the coupling constant becomes larger and larger, or the energy scale
is decreased to the low-temperature regime ($T=T_c$). 

However, in the large-$\bar{U}$ limit, the Schrieffer-Wolff canonical 
transformation can also be applied to the present model, and a s-d type of 
model (so-called compactified two-channel Kondo model \cite{cit}) is obtained
\begin{equation}
H=it\sum_n\sum_{\alpha=0}^3\Psi_{\alpha}(n+1)\Psi_{\alpha}(n)
  +J \left [\vec{\sigma}(0)+\vec{\tau}(0)\right] \cdot \vec{S}_d,
\end{equation}
where $J=2V^2/U$ and $\rho J=\frac{2}{\pi^2}\left(\frac{\pi\Delta}{U}\right)$.
So the weak-coupling limit $\rho J\ll 1$ of this s-d model corresponds to the
large-$\bar{U}$ limit, and the multiplicative RG analysis can be also used to
this model as the usual treatments for the ordinary Kondo model 
\cite{amfz,solyom}. The scaling equation has been derived as
\begin{equation}
 \frac{d(\rho J)}{d\ln D}=-[2(\rho J)^2-(\rho J)^3],
\end{equation}
where the conduction electron bandwidth $D$ plays the role of the high-energy 
cut-off. This scaling equation explicitly shows that the dimensionless coupling
parameter $\rho J$ grows stronger and stronger under the RG transformations. 
For the corresponding compactified Anderson model in the large $\bar{U}$ 
regime, this is equivalent to a decrease in the dimensionless coupling 
parameter $\bar{U}$ ($\rho J=2/(\pi^2\bar{U})$) as the low-energy scale is 
reduced. 
Using the relevant cut-off factors in the small and large $\bar{U}$ regime, we
arrive at a schematic flow diagram as shown in Fig.6,
and we conjecture a stable fixed point of the model Hamiltonian
in the intermediate coupling regime at $\bar{U}=\bar{U}_c$. This flow diagram 
is analogous to that of the isotropic two-channel Kondo model \cite{nb}. 
Obviously, there are significant differences between the present compactified 
Anderson impurity model and the two-channel Kondo model. 

In conclusion, we have developed a systematic perturbation theory for the 
compactified Anderson impurity model in the Majorana fermion representation,
where the unperturbed Hamiltonian has a degenerate ground state. We have 
calculated the leading perturbational corrections in the
weak coupling limit. We have also derived some general relations for the
impurity susceptibilities and self energies in terms of the vertex functions.
The main results of the paper are that a linear temperature dependence of the 
electrical resistivity is obtained from the second order theory, and some non-FL
thermodynamic properties have been calculated as well. The conduction electron 
singlet superconducting pairing correlation function at the impurity site is 
found singular in second order in $U$, indicating the formation of the 
singlet conduction electron pairing resonance at the Fermi level. In the third 
order in $U$, the vertex function $\Gamma_{0,1,2,3}(0,0,0,0)$ has 
logarithmic corrections, and the summation of the leading order logarithmic
terms results in a new weak-coupling low-temperature energy scale 
$T_c
=\Delta{\rm exp}\left[-\frac{1}{9}\left(\frac{\pi\Delta}{U}\right)^2\right]$,
below which the perturbational approach begins to break down. 
The behavior of the low-energy excitations below $T_c$ is still open for the 
future investigations. 

We are grateful to the SERC for the support of a research grant. G. M. Zhang 
would like to thank Prof. Lu Yu and Prof. Zhao-Bin Su for the early 
collaborations on impurity problems.

\newpage
\begin{figure}
\caption {The Dyson's equations of the impurity self energies in terms of the
impurity vertex function $\Gamma_{0,1,2,3}$. (a) is for $\Sigma_{\rm sc}$, and
(b) is for $\Sigma_{\rm vec}$. The solid lines denote the perturbed propagators
$G_{\rm vec}$ and the dotted lines are $G_{\rm sc}$.}
\label{fig1}
\end{figure}

\begin{figure}
\caption {The even and odd parts of the impurity spin and charge dynamical
susceptibilities in terms of the impurity propagators and vertex 
functions. The thick solid lines denote the perturbed propagators 
$G_{\rm vec}$ and the thick dotted lines are $G_{\rm sc}$, while 
the thin solid lines correspond to the unperturbed propagators $G_{\alpha}$ 
and the thin dotted lines to $G_0$. }
\label{fig2}
\end{figure}

\begin{figure}
\caption{The second-order perturbative corrections to the impurity self 
energies. (a) is for the scalar field, and (b) for the vector field.}
\label{fig3}
\end{figure}

\begin{figure}
\caption {The first order and third order corrections to the impurity vertex
function $\Gamma'_{0,1,2,3}(0,0,0,0)$. (a) is the first order correction. (b),
(c), (d), and (e) are the third order contributions. But only the diagram (b)
gives the logarithmically temperature dependent correction.}
\label{fig4}
\end{figure}

\begin{figure}
\caption {The perturbative corrections to the even and odd part of the 
impurity spin and charge static susceptibilities in the low order 
perturbations. (a) is the singular diagram of the even part in the zero order 
in $U$. (b) is the odd part in the first order
perturbations. (c) is the $\ln T$ contribution to the even part in the
second-order of $U$. (d) is the diagram giving rise to the $\ln^2 T$ terms of
the even part in the second-order perturbations. }
\label{fig5}
\end{figure}

\begin{figure}
\caption
{The schematic RG flow diagram in terms of the dimensionless coupling parameter
$\bar{U}=U/(\pi\Delta)$. The arrows denote the direction of decreasing the 
relevant high-energy scale. $\bar{U}_c$ corresponds to the stable fixed point 
of the model.}
\label{fig6}
\end{figure}


\newpage
\begin{references}
\bibitem{maple} For reviews, see, e.g.,
M. B. Maple, C. L. Seaman, D. A. Gajewski, Y. Dalichaouch, V. B. Barbetta,
M. C. de Andrade, H. A. Mook, H. G. Lukefahr, O. O. Bernal, D. E. MacLaughlin,
J. Low Temp. Phys. {\bf 95}, 225 (1994);
H. L\"{o}hneysen, Physica B {\bf 206-207}, 101 (1995).

\bibitem{varma}C. M. Varma, P. B. Littlewood, S. Schmitt-Rink, E. Abrahams, and
A. E. Ruckenstein, Phys. Rev. Lett. {\bf 63}, 1996 (1989);
 C. M. Varma, Int. J. Mod. Phys. B {\bf 3}, 2083 (1989).

\bibitem{and}P. W. Anderson, Phys. Rev. Lett. {\bf 64}, 1839 (1990). 
 
\bibitem{hewson}For a general review, see A. C. Hewson, 
 {\it The Kondo Problem and Heavy Fermions} (Cambridge University Press, 1992).

\bibitem{nb}P. Nozi\`{e}res and A. Blandin, J. Phys. {\bf 41}, 193 (1980).

\bibitem{adtw}N. Andrei, and C. Destri, Phys. Rev. Lett. {\bf 52}, 364
(1984); A. M. Tsvelik and P. B. Wiegmann, J. Stat. Phys. {\bf 38}, 125
(1984).

\bibitem{al}I. Affleck and A. W. W. Ludwig, Phys. Rev. Lett. {\bf 67},
3160 (1991); Nucl. Phys. B {\bf 360}, 641 (1991); 
Phys. Rev. B {\bf 48}, 7279 (1993).

\bibitem{ek} V. J. Emery and S. Kivelson, Phys. Rev. B {\bf 46}, 10812
 (1992); A. M. Sengupta and A. Georges, {\it ibid.}, {\bf 49}, 10020 (1994);
D.G. Clarke, T. Giamarchi, and B. I. Shraiman $ibid$, {\bf 48}, 7070 (1993).

\bibitem{cit}P. Coleman, L. Ioffe, and A. M. Tsvelik, Phys. Rev. B {\bf 52},
6611 (1995).

\bibitem{cs}P. Coleman and A. J. Schofield, Phys. Rev. Lett. {\bf 75},
2184 (1995).

\bibitem{zhang}Guang-Ming Zhang and A. C. Hewson, Phys. Rev. Lett. {\bf 76},
2137 (1996). 

\bibitem{hews}A. C. Hewson, in preparation.

\bibitem{gh}For the properties of Pfaffian determinant, see
H. S. Green and C. A. Hurst, {\it Order-Disorder Phenomena}
(Interscience Publishers, 1964).

\bibitem{yy}K. Yosida and K. Yamada, Prog. Theor. Phys. {\bf 46}, 244 (1970);
$ibid$, {\bf 53}, 1286 (1975).

\bibitem{yamada}K. Yamada, Prog. Theor. Phys. {\bf 53}, 970 (1975); $ibid$, 
{\bf 54}, 316 (1975).

\bibitem{bs}N. N. Bogoliubov and D. V. Shirkov,
{\it Introduction to the Theory of Quantized Fields}
(London, Interscience, 1959).

\bibitem{amfz}A. A. Abrikosov and A. A. Migdal, J. Low Temp. Phys. {\bf 3},
 519 (1970); M. Folwer and A. Zawadowski, Solid State Commun. {\bf 9},
 471 (1971).

\bibitem{solyom}J. S\'{o}lyom, J. Phys. F {\bf 4}, 2269 (1974).

\bibitem{anderson}P. W. Anderson, J. Phys. C {\bf 3}, 2436 (1970);
 P. W. Anderson, G. Yuval, and D. R. Hanman, Phys. Rev. B {\bf 1}, 4464 (1970).

\bibitem{wilson}H. R. Krishna-murthy, J. W. Wilkins, and K. G. Wilson,
Phys. Rev. B {\bf 21} 1008 (1980); $ibid$, {\bf 21}, 1044 (1980).

\bibitem{haldane}F. D. M. Haldane, J. Phys. C {\bf 14}, 2585 (1981); 
Phys. Rev. Lett. {\bf 40}, 416 (1978).
  
\end{references}
\end{document}